\documentclass[aps,preprint,onecolumn,floatfix,superscriptaddress,nofootinbib]{revtex4-1}
\usepackage[T1]{fontenc} 
\usepackage{multirow}
\usepackage{makecell}
\usepackage{longtable}

\newcommand{\sigsip}{\ensuremath{\sigma^{\rm SI}_{\chi p}}}

\newcommand{\sv}{\ensuremath{\langle\sigma v\rangle}}
\newcommand{\dnde}{\ensuremath{\frac{dN_e}{dE_e}}}
\newcommand{\dndEga}{\ensuremath{\frac{dN_\gamma}{dE_\gamma}}}

\newcommand{\gev}{\ensuremath{\,\mathrm{GeV}}}

\usepackage{amsmath}
\usepackage{graphicx,bm}
\usepackage[colorlinks,citecolor=blue]{hyperref}
\usepackage[caption=false]{subfig}
\usepackage{slashed}
\usepackage{ulem}

\begin{document}
\title{Muonphilic Dark Matter explanation of gamma-ray galactic center excess: a comprehensive analysis}

\author{Murat Abdughani}
\affiliation{School of Physical Science and Technology, Xinjiang University,Urumqi 830046, China}
\affiliation{Key Laboratory of Dark Matter and Space Astronomy, Purple Mountain Observatory, Chinese Academy of Sciences, Nanjing 210033, China}

\author{Yi-Zhong Fan}
\affiliation{Key Laboratory of Dark Matter and Space Astronomy, Purple Mountain Observatory, Chinese Academy of Sciences, Nanjing 210033, China}
\affiliation{School of Astronomy and Space Science, University of Science and Technology of China, Hefei, Anhui 230026, China}

\author{\\Chih-Ting Lu}
\affiliation{Department of Physics and Institute of Theoretical Physics, Nanjing Normal University, Nanjing, 210023, China}

\author{Tian-Peng Tang}
\affiliation{Key Laboratory of Dark Matter and Space Astronomy, Purple Mountain Observatory, Chinese Academy of Sciences, Nanjing 210033, China}
\affiliation{School of Astronomy and Space Science, University of Science and Technology of China, Hefei, Anhui 230026, China}

\author{Yue-Lin Sming Tsai}
\affiliation{Key Laboratory of Dark Matter and Space Astronomy, Purple Mountain Observatory, Chinese Academy of Sciences, Nanjing 210033, China}

\begin{abstract}
The Galactic center gamma-ray excess (GCE) is a long-standing unsolved problem. One of candidate solutions, the dark matter (DM) annihilation, has been recently tested with other astrophysical observations, 
such as AMS-02 electron-positron spectra, Fermi Dwarf spheroidal galaxies gamma-ray data, and so on. 
By assuming that the DM particles annihilate purely into a normal charged fermion pair, 
Di Mauro and Winkle (2021) claimed that 
only a muon-pair is compatible with the null detection of all the corresponding astrophysical measurements and can explain GCE simultaneously. 
On the other hand, a muonphilic DM model may also lead 
to a signal in the recent Fermilab muon $g-2$ measurement or be constrained by the latest PandaX-4T limit. 
In this work, we comprehensively study interactions between DM and muon, 
including various combinations of DM and mediator spins.
In agreement with GCE (not only $2\mu$ but also $4\mu$ final states), we test these interactions  
against all the thermal DM constraints. Our results show that only the parameter space 
near the resonance region of mediator can explain GCE and relic density simultaneously, 
and larger parameter spaces are still allowed if other poorly-known systematic uncertainties are included. 
Regardless of the DM spin, only the interactions with the spin-0 mediator can explain 
the recent muon $g-2$ excess on top of GCE, relic density, and other DM and mediator constraints.
\end{abstract}

\date{\today}

\maketitle

\tableofcontents

\newpage

\section{Introduction} 

Dark matter (DM) is a successful candidate to consistently explain many astrophysical and cosmological problems. 
Except for those known gravitational DM evidence, 
we are still seeking for any non-gravitational interaction between 
DM and the visible matter in order to pin-down the DM particle nature by means of collider experiments~\cite{ATLAS:2021shl,CMS:2021far}, 
DM direct detection (DD)~\cite{XENON:2018voc,PandaX-4T:2021bab} and indirect detection (ID)~\cite{Fermi-LAT:2017bpc,DAMPE:2017fbg,AMS:2021nhj}. 
Among those non-gravitational detection, the Galactic center gamma-ray excess (GCE) reveals a possibility
that the DM annihilation with the mass around $30-70\gev$ in the Galactic center 
can well fit the shapes of the energy and spatial spectra~~\cite{Hooper:2010mq,Zhou:2014lva,Calore:2014xka,Daylan:2014rsa,2101.04694}. 
However, the origin of this GCE has been a long-standing controversy. 
One possible astrophysical explanation is that some undetected millisecond pulsars are concentrated near the center of the Milky Way and in the bulge throughout the Galaxy \cite{1904.08430,2002.12370,1611.06644,1711.04778,1901.03822}. Another explanation involves an outburst of leptons or hadrons accelerated by the supermassive black hole, known as the Sgr $\rm A^*$ \cite{Petrovic:2014uda, Cholis:2015dea, Carlson:2014cwa}.

The systematic uncertainties of these GCE analyses are still unclear. It can be a challenge to discover or exclude
the DM origin by only using GCE Fermi data.  
A strategy to test the DM origin is to cross-check against other astrophysical data, 
such as Fermi-LAT observations of dwarf spheroidal galaxies (dSphs; \cite{Li:2021vqg}) and AMS-02 cosmic-ray data \cite{AMS:2021nhj}.
Once all the above data do not support DM annihilation, we may abandon the DM explanation of GCE.  
Motivated by such a consideration, a recent work~\cite{DiMauro:2021qcf} has performed a combined analysis by taking  
the $\gamma$-ray data of 48 dSphs and the latest AMS-02 positron and antiproton data into account. 
They focus on the DM annihilation to a pair of the Standard Model charged fermion $f$ final state, namely ${\rm DM+DM} \to f+\bar{f}$. 
When the final state particle mass is heavier than 
$80\gev$ (for instance $ZZ$, $WW$, $HH$, and $t\bar{t}$),  
the DM annihilation cannot generate a well-fit spectrum to GCE data. 
Although the pure hadronic or some mixture between leptonic and hadronic final states of DM annihilation can fit GCE data, 
their required vertical half-height of the diffusive zone for the AMS-02 antiproton data ($L\simeq 2$~kpc)   
are  in tension with the fitted value of  
the radioactive cosmic ray and radio data ($L=4.1^{+1.3}_{-0.8}$~kpc)~\cite{Weinrich:2020ftb} at a confidence level of  $\sim 2-3\sigma$. 
On the other hand, the final state $e^+e^-$ of DM annihilation would be disfavoured by the AMS-02 $e^+$ data 
where the authors adopted semi-analytic propagation equation and the uncertainties from $L$ and 
DM halo profiles were addressed. 
Finally, Ref.~\cite{DiMauro:2021qcf} claims that DM annihilation into the muon final state (called muonphilic DM hereafter) 
can reasonably explain GCE without violating other astrophysical constraints\footnote{Note that the authors in~\cite{DiMauro:2021qcf} did not consider 
the mixture of charged fermion pairs and neutrino pairs in the final state. 
Since the final state neutrino does not affect the gamma-ray spectrum, 
this situation is somehow similar to the pure muon-pair case.}. 
However, the astrophysical systematic uncertainties (both from propagation equation and unknown sources) 
involved in~\cite{DiMauro:2021qcf} could have been underestimated, see e.g.~\cite{Calore:2014xka}. 
Thus, to further probe such a muonphilic DM model, implementing the explicit interaction terms 
confronting with other data are highly required.

We propose to verify muonphilic DM GCE signal by using the particle experimental constraints.    
There are at least four important motivations.
First, the latest DM DD limit given by PandaX-4T~\cite{PandaX-4T:2021bab} provides 
a severe constraint for DM and nucleon scattering. 
If DM would only couple to muon, it naturally generates a loop-suppressed DM-nucleon scattering cross section.  
Hence, it is not surprising that the muonphilic DM can explain the GCE and escape the constraint from the PandaX-4T detection. 
Second, the most recently reported excess of the muon $g-2$ measurement by the FermiLab E989 experiment is  
$\delta a_{\mu}= (2.51 \pm 0.59) \times 10^{-9}$, which deviates from the standard model prediction at a confidence level of $4.2\sigma$~\cite{Muong-2:2021ojo}. 
Although the sign of $\delta a_{\mu}$ can be either positive or negative depending on the mediator (MED) nature,
the combined result can restrict the parameter space of the muonphilic DM models.  
Third, the relic density measurement with the thermal DM paradigm can further narrow down the parameter space. 
The interplay between the annihilation cross sections at the early and present time can be highly non-trivial. 
Conventionally, the annihilation cross section can be simply expanded by the power of relative velocity, 
namely $\sv\simeq a+b v_{\rm rel.}^2$ with dropping the higher order contribution. 
In the partial wave approach, one can define that the $s$-wave contribution is from $a$ while the $p$-wave contribution is from $b v_{\rm rel.}^2$. 
Thus, the relative velocities in the early universe for the relic density and the present universe for the GCE are very different. 
It is interesting to check whether the muonphilic DM explanation to the GCE is supported by the PLANCK relic density measurement. 
Finally, the muonphilic DM models with $Z_2$-even mediators can easily escape the mono-photon and mono-jet constraints from LEP~\cite{Fox:2011fx} and LHC~\cite{ATLAS:2021kxv,CMS:2021far} such that the electroweak scale DM is still allowed. 
Furthermore, the future muon colliders can be used to test these muonphilic DM models~\cite{Huang:2021nkl,Capdevilla:2021rwo,AlAli:2021let}. For the $Z_2$-even MED, $\mu^+\mu^-\rightarrow\mu^+\mu^-$ process is powerful to directly search for the MED and the mono-$\gamma$ process can be used to explore DM when a DM pair is produced form the on-shell MED. On the other hand, for the $Z_2$-odd MED, the MED can be searched for via its pair production process and the mono-$\gamma$ process from $t$-channel DM pair production is again used to explore DM.

In this work, we comprehensively list all the possible renormalizable interactions by simply appending a DM and a MED to 
the standard model (SM). We restrict ourselves to only concern SM singlet DM and MED 
with the spins ($s=0,1/2,1$). In total, we have 16 interaction types for $Z_2$-even mediator while 
7 interaction types for $Z_2$-odd mediator. We will investigate all these 23 interaction types and eliminate some disfavoured ones 
by using a global analysis with the likelihoods from PLANCK relic density~\cite{Planck:2015fie}, Fermi GCE~\cite{DiMauro:2021raz}, PandaX-4T limits~\cite{2107.13438}, 
the LEP limit~\cite{ALEPH:2002gap}, and $\delta a_{\mu}$~\cite{Muong-2:2021ojo}.

The remainder of this paper is structured as follows. 
In Sec.~\ref{sec:4mu}, we recap the explanation of the GCE by using the DM annihilation to $2\mu$ scenario. 
Additionally, we include $4\mu$ final state that can also mimic the signature of $2\mu$ final state. 
In Sec.~\ref{sec:likelihood}, we summarize all the relevant experimental likelihoods used in our numerical work. 
After a comprehensive discussion of all the possible interaction types in Sec.~\ref{sec:models}, 
we can eliminate several disfavoured ones.
In Sec.~\ref{sec:dd} and~\ref{sec:g2}, we further evaluate the future detectability of the DD and muon $g-2$ experiments, respectively. 
Finally, we summarize and conclude our results in Sec.~\ref{sec:conclusions}. 
Some detailed formulas for calculations are included in three appendices. 

\section{The muonphilic DM explanation to the GCE} \label{sec:4mu}

It is claimed in~\cite{DiMauro:2021qcf} that 
all hadronic and semi-hadronic annihilation channels 
can be excluded by the AMS-02 antiproton data, unless 
the height of the diffusion halo $z_h$ is smaller than $2$~kpc which is however in tension with the radio data. 
Except for $\mu^+\mu^-$ final state, these authors also found that DM annihilation to the leptonic channels 
can be ruled out by either the combined dSphs limits or AMS02 positron data. 
Therefore, it is concluded in~\cite{DiMauro:2021qcf} that the DM annihilation to a pair of muons with the mass around $60~\gev$, 
decaying to electrons subsequently, can explain GCE via inverse Compton scattering (ICS) with starlight. 
The DM prompt $\gamma$ emission, mainly from final state radiation, 
can also contribute to the gamma-ray fluxes at the higher energy range. 
The propagation of $e^\pm$, gamma-ray emission of ICS and prompt $\gamma$ at the GC are 
summarized in Appendix~\ref{sec:ICS}. 
The favoured annihilation cross sections ($\mu^+\mu^-$ final state) and DM masses are~\cite{DiMauro:2021qcf} 
\begin{eqnarray}
\sv_{2\mu}= 3.9^{+0.5}_{-0.6} \times 10^{-26}~{\rm cm}^3 s^{-1}, ~~{\rm and}~m_D =58^{+11}_{-9} \gev.
\label{eq:gce_data}
\end{eqnarray}

The muonphilic DM can annihilate into a pair of light mediators at the present time. 
This annihilation to a pair of light mediators leads four muons in the final state 
and its spectrum can differ from the one of the $2\mu$ final state. 
The electron energy spectrum generated from the $4\mu$ final state can be written as 
\begin{eqnarray}
\dnde^{[4\mu]}(m_D, M, E_e)=\int_{E_{\rm min}}^{E_{\rm max}} dE_\mu \frac{dN_\mu}{dE_\mu}(m_D, M, E_\mu) 
\dnde^{[2\mu]}(E_\mu, E_e), 
\label{eq:dnde}
\end{eqnarray}
where $M$ is the mass of the mediator.  
The energy of electron and muon are $E_e$ and $E_\mu$, respectively. 
We can take the spectrum $\dnde^{[2\mu]}(E_\mu, E_e)$ from \texttt{PPPC4}~\cite{1012.4515,1009.0224} 
by using the central energy equal to $2 E_\mu$ instead of $2 m_D$, while  
$\frac{dN_\mu}{dE_\mu}(m_D, M, E_\mu)$ is a box-shape spectrum, 
\begin{eqnarray}
\frac{dN_\mu}{dE_\mu}(m_D, M, E_\mu)= \frac{4}{E_{\rm max}-E_{\rm min}} 
\Theta(E_\mu-E_{\rm min})\Theta(E_{\rm max}-E_\mu), 
\label{eq:box}
\end{eqnarray}
where $\Theta$ is the Heaviside function. The maximum and minimum muon energy are 
\begin{eqnarray}
 E_{\rm max/min}&=& \frac{m_D}{2} \times \left(1\pm\sqrt{1-\frac{M^2}{m_D^2}} \right). 
\label{eq:Emuon}
\end{eqnarray}
Similarly, we can replace $\dnde^{[2\mu]}(E_\mu, E_e)$ with $\dndEga^{[2\mu]}(E_\mu, E_\gamma)$ 
to obtain the DM prompt gamma-ray contribution for $4\mu$ case.

\begin{figure}
\includegraphics[width=1.1\textwidth]{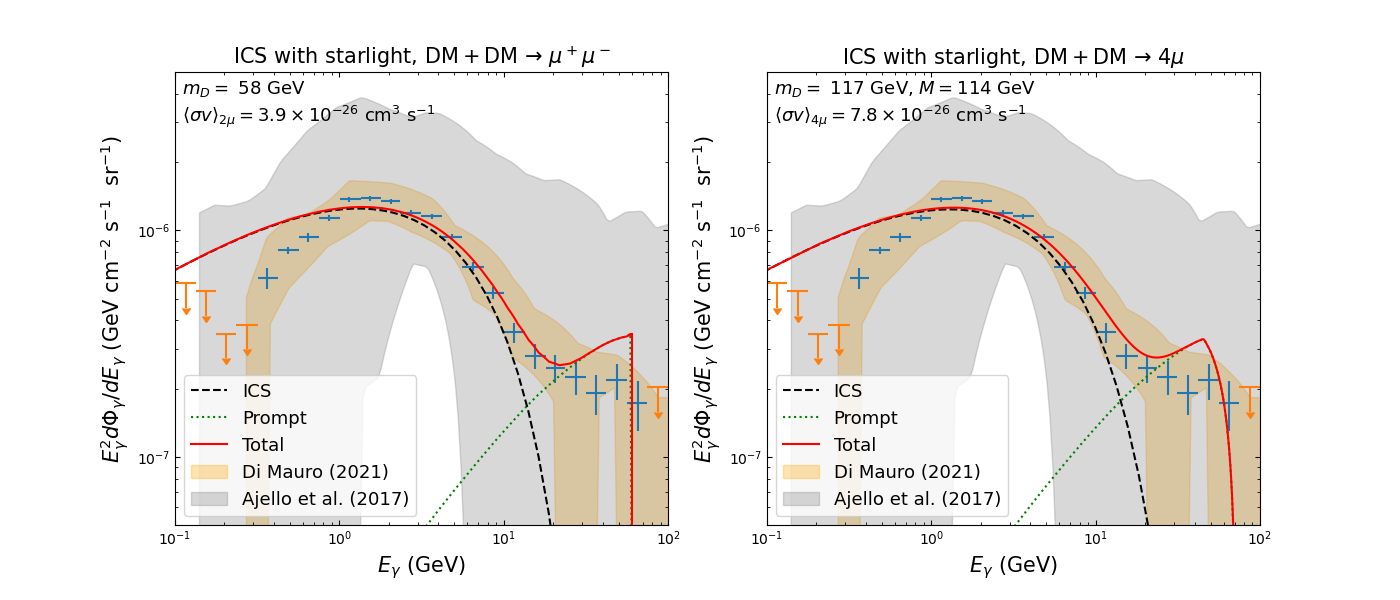}
\caption{The gamma-ray flux component for ICS (black dashed line) 
and prompt emission (green dotted line) from DM annihilation to $\mu^+ \mu^-$ (left panel) 
and $4\mu$ (right panel) final states. 
The combination of ICS and prompt emission is represented by red solid line. 
The DM mass $m_D$, mediator mass $M$, and annihilation cross sections $\langle \sigma v\rangle_{2\mu / 4\mu}$ are best-fit DM parameters. 
While the blue and the orange GCE data points are reported in Ref.~\cite{2101.04694}, 
only blue data points are considered in our $\chi^2$ calculations.}
\label{fig:sv2sv4}
\end{figure}

By plugging Eq.~\eqref{eq:dnde} into \texttt{RX-DMFIT}~\cite{1705.09384}, we can compute the 
fluxes of DM induced ICS with starlight. 
However, the fluxes of the DM prompt $\gamma$ emission can be directly computed by using Eq.~\eqref{eq:promt}. 
As implemented in \texttt{RX-DMFIT}, we take the diffusion zone radius of $r_h = 30$~kpc 
and the distance of the sun from galaxy center is $r_\odot = 8.33$~kpc. 
For the diffusion coefficient model as demonstrated in Appendix~\ref{sec:ICS}, 
we adopt the representative galactic diffusion parameter values\footnote{It has been shown 
in Ref.~\cite{1705.09384} that the propagation uncertainties of 
DM ICS spectra are rather small compared with the uncertainties from the background.}, 
i.e., $D_0 = 3 \times 10^{28}$ cm$^2\, s^{-1}$ and $\gamma = 0.3$~\cite{Webber:1992dks,Baltz:1998xv,Strong:2007nh,Vladimirov:2011rn,McDaniel:2018vam,Salati:2021hnj}. 
In addition, we take ``non-cool-core'' magnetic field model
\begin{equation}
    B(r) = B_0 e^{-r/r_c}, 
\end{equation}
with the core radius of $r_c=3$~kpc, the central magnetic field of $B_0=4.7~\mu$G. 
For the DM density distribution, we choose Navarro-Frenk-White (NFW) profile~\cite{astro-ph/9508025,astro-ph/9611107}
\begin{equation}
    \rho (r) = \frac{\rho_s}{\left( \frac{r}{r_s} \right) \left( 1+ \frac{r}{r_s} \right)^2},
\end{equation}
with the characteristic density $\rho_s = 0.184\gev/$cm$^3$ and radius $r_s = 24.42$~kpc~\cite{1505.01049}. 
Following Ref.~\cite{astro-ph/0507575}, 
the parameter values for the energy loss coefficients in units of $10^{-16}\gev/s$ are taken 
to be $b^0_{\rm syn} \simeq 0.0254$, $b^0_{\rm IC} \simeq 0.25$, $b^0_{\rm brem} \simeq 1.51$, 
$b^0_{\rm Coul} \simeq 6.13$, and $b_{\rm ICSL} = 6.08$ for Milky Way galaxy model~\cite{1001.4086}. 
The average thermal electron number density is taken as $n_e \approx 0.1~{\rm cm^{-3}}$~\cite{astro-ph/0702532}. 
We note that only $b_{\rm ICSL}$ among the energy loss terms is sensitive to our conclusion.

In Fig.~\ref{fig:sv2sv4}, we present the gamma-ray fluxes produced for ICS (black dashed lines) and prompt (green dotted lines) emission from DM particles annihilating 
into $\mu^+ \mu^-$ (left panel) and $4\mu$ (right panel) final states, as well as their combinations (red solid lines). 
We choose the central value of DM parameters in Eq.~\eqref{eq:gce_data} for the $2\mu$ final state 
while we evaluate the best-fit parameters for $4\mu$ final state.
By comparing the fluxes of $2\mu$ and $4\mu$ final states, 
despite some differences between their electron spectra, 
they can have similar shapes after convolution with the starlight photon density. 
The Eq.~\eqref{eq:dnde} infers that the spectral shape of $\dnde^{[4\mu]}$ is basically the same as $\dnde^{[2\mu]}$ with $m_D\simeq M/2$, 
but the former is larger than the latter by a factor of 2. 
Thus, if requiring the same ICS gamma-ray fluxes to explain GCE, 
a twice higher annihilation cross section is needed, see the right panel of Fig.~\ref{fig:sv2sv4}. 
Therefore, it will be difficult to explain the GCE and relic density measurement simultaneously 
in the sceanrio of ${\rm DM}+{\rm DM}\to{\rm MED}+{\rm MED}$.

\section{The likelihoods}
\label{sec:likelihood}

In this work, we mainly consider three important likelihoods. 
Although Fermi GCE and PLANCK relic density are based on the signal, 
the DM direct detection from PandaX-4T can set an upper limit on the interaction. 
In the below, we will present their $\chi^2$ and the total $\chi^2_{\rm tot}$ defined as the sum of 
individual $\chi^2$ values of GCE, DM relic density, and DD cross section 
\begin{equation}
    \chi^2_{\rm tot} = \chi^2_{\rm GCE} + \chi^2_{\Omega h^2} + \chi^2_{\rm DD}.
\end{equation} 
We hire \texttt{emcee}~\cite{Foreman-Mackey:2012any} based on Markov Chain Monte Carlo (MCMC) method 
to undertake the task of sampling the parameter space 
with the likelihood $\propto\exp(-\chi^2_{\rm tot}/2)$. 
We use \texttt{Feynrules}~\cite{1310.1921} to implement the models, and then import them 
to \texttt{MicroMEGAS}~\cite{1004.1092} for DM relic density calculation. 
The number of samples for each model in 2$\sigma$ and 3$\sigma$ ranges are about $3\times10^5$ and $4.5\times10^5$ respectively.

\begin{itemize}
    \item \textbf{\underline{Fermi GCE}}:\\
We accommodate the GCE reduced $\chi^2$ as
\begin{equation}
    \chi^2_{\rm GCE} = \sum_{i=1}^{19} \left( \frac{dN}{dE_i} - \frac{dN_0}{dE_i} \right)^2 / 19 \sigma_i^2,
    \label{eq:GCE}
\end{equation}
where $\frac{dN}{dE_i}$, $\frac{dN_0}{dE_i}$ and $\sigma_i$ are predicted gamma-ray spectra, GCE spectra extracted from
Fermi-LAT data after background modeling and their errors~\cite{2101.04694}. 
Here, we simply ignore the orange error bars and systematic uncertainties (gray and orange bands) in Fig.~\ref{fig:sv2sv4} 
as well as the correlation between energy bins.
Therefore, the total number of data bins used for our analysis (blue error bars in Fig.~\ref{fig:sv2sv4}) are 19.   
The predicted gamma-ray spectrum is
\begin{equation}
    \frac{dN}{dE} = \frac{dN}{dE}^{[2\mu]} \times {\rm BR}_{2\mu} + \frac{dN}{dE}^{[4\mu]} \times (1-{\rm BR}_{2\mu}), 
\end{equation}
where the annihilation fraction ${\rm BR}_{2\mu}$ describes the portion of the $2\mu$ annihilation final state. 

In Ref.~\cite{DiMauro:2021qcf}, they obtained a minimum reduced $\chi^2$ of 5.47 for the $2\mu$ final state. 
This quoted value is not located at around one because some uncertainties 
such as the model uncertainties of the Galactic gas and the interstellar radiation field  
are not taken into account\footnote{We gratefully acknowledge the private communication 
with Mattia Di Mauro, the author of Ref.~\cite{DiMauro:2021qcf}.}. 
However, it is unable to reach a consensus about the precise uncertainties, e.g., 
the gray and orange bands in Fig.~\ref{fig:sv2sv4}. 
When we ignore these poorly-known systematic uncertainties, 
a minimum reduced $\chi_{\rm red}^2$ of 5.15 (4.34) for the $2\mu$ ($4\mu$) final state are obtained. 
However, the fact that the reduced $\chi^2$ is not more or less equal to 1 implies that 
some systematic uncertainties may be overlooked.  
Therefore, we adopt two statistical approaches to demonstrate our results in two colors in our plots. 
In the first approach (green layer in all the scatter plots), 
according to Ref.~\cite{ParticleDataGroup:2020ssz} (see Sec. 5.2 there), 
if the minimum reduced chi-square $\chi_{\rm red}^2$ is larger but not much greater than 1, 
then we can enlarge our errorbars with square-root of the minimal reduced $\chi_{\rm red}^2$, namely 
\begin{equation}
\sigma_{i,{\rm GCE}}^\prime=\sqrt{\chi_{\rm red}^2}\times \sigma_i, 
\end{equation}
where $\sigma_{i,{\rm GCE}}^\prime$ is used in our study but $\sigma_i$ is the same as defined in Eq.~\eqref{eq:GCE} of  Ref.~\cite{2101.04694}. 
Apparently, the new overall GCE reduced chi-squares are smaller 
and the new minimum value of them is exactly equal to 1. 
Thus, our analysis to examine the allowed parameter space is more conservative than 
the one directly computed by Eq.~\eqref{eq:GCE}.   
In the second approach (gray layer in all the figures), we will also perform a comparison with  
the allowed spectra in agreement with the gray region as shown in Fig.~\ref{fig:sv2sv4}.  
For the most conservative way, the second approach allows us to project the systematic uncertainties 
from propagation and source models to our particle model parameter space.

\item \textbf{\underline{PLANCK relic density}}:\\
The DM PLANCK relic density $\chi^2$ is described as a Gaussian distribution 
\begin{equation}
    \chi^2_{\Omega h^2} = \left( \frac{\mu_t - \mu_0}{\sqrt{\sigma_{\rm theo}^2+\sigma_{\rm exp}^2}} \right)^2,
\end{equation}
where $\mu_t$ is predicted from the theoretical value, $\mu_0$ is an experimental central value, and theoretical uncertainty $\sigma_{\rm theo} = \tau \mu_t$. 
We use PLANCK 2018 data~\cite{Planck:2015fie} to constrain our predicted relic density $\Omega h^2$. 
Their reported central value with statistical error is $\Omega h^2 = 0.1186 \pm 0.002$. 
On the other hand, we may also need to address the uncertainties from the Boltzmann equation solver and the entropy table in the early universe. 
Hence, we conservatively introduce $\tau = 10\%$ based on our prediction.

\item \textbf{\underline{PandaX-4T $\sigsip$}}:\\
The estimation of $\chi^2$ for the DM-nucleus spin-independent (SI) direct detection cross section $\chi^2_{\rm DD}$ is
\begin{equation}
    \chi^2_{\rm DD} = \left( \frac{\sigma^{\rm SI}_{\chi p}}{\sigma^{\rm SI, 90\%}_{\chi p}/1.64} \right)^2,
\end{equation}
where $\sigma^{\rm SI}_{\chi p}$ and $\sigma^{\rm SI, 90\%}_{\chi p}$ are predicted from 
the theoretical value and upper limits of the cross sections for a given DM mass at $90\%$ confidence level 
from PandaX-4T~\cite{2107.13438}, respectively. 
By assuming null detection, we can take the central value as zero and 
the number $1.64$ is the unit of $90\%$ confidence level.   
\end{itemize}

\section{Possible interaction types}
\label{sec:models}

\begin{table}[h]
\centering
\begin{tabular}{|c|c|c|c|}
\hline
 & Scalar & Fermion & Vector \\
\hline \hline
Dark Matter & $S$ & $\chi$ & $X^{\mu}$ \\
\hline
Mediator & $\phi$ & $\psi$ & $V^{\mu}$ \\
\hline
\end{tabular}
\caption{The particle notation for DM and MED with various spin types used in our work.}
\label{tab:particle_notataion}
\end{table}

In this section, we summarize all representative interaction types for two DM particles annihilating into 
a pair of muons and a pair of MEDs at tree level. 
By taking a $Z_2$ symmetry to prevent DM decay, we introduce $Z_2$-even 
mediators for s-channel while $Z_2$-odd mediators for t-channel annihilation. 
As presented in Table~\ref{tab:particle_notataion}, 
both DM and MED can be scalar (spin-0), fermionic (spin-$\frac{1}{2}$), or vector (spin-1). 
In this work, we will discuss self-conjugate and not self-conjugate DM fields, \textit{i.e.},  
(i) real and complex scalar DM, (ii) Majorana and Dirac DM, and (iii) real and complex vector DM. 
For the sake of simplicity, we only concern a self-conjugate field for $Z_2$-even mediator in this study.
However, a complex field is required for $Z_2$-odd mediator.

\subsection{$Z_2$-even mediator}

\begin{table}[t]
    \centering
    \begin{tabular}{c|c|c|c|c}
\hline\hline
\multicolumn{5}{c}{$Z_2$ even mediator} \\
\hline \hline
 types  & Lagrangian & $\langle \sigma v\rangle_{2\mu}$ & $\langle \sigma v\rangle_{4\mu}$ & DD  \\
        &            & $\simeq a+b v^2$ & $\simeq a+b v^2$ &   \\
\hline \hline
\multirow{4}{*}{$\chi$ and $\phi$}  & $\mathcal{L}_1=(g_{D} \bar \chi \chi + g_f \bar f f )\phi$
& $a=0$ & $a=0$ & Eq.~\eqref{eq:DDL1}  \\

    & $\mathcal{L}_2=(g_D\bar \chi \chi + g_f \bar f i \gamma^5 f)\phi$ & $a=0$ & $a=0$ & ---  \\

    & $\mathcal{L}_3=(g_D\bar \chi i \gamma^5 \chi + g_f \bar f f)\phi$ & \textbf{Case (i)} & $a=0$ & Eq.~\eqref{eq:DDL3}  \\

    & $\mathcal{L}_4=(g_D\bar \chi i \gamma^5 \chi + g_f \bar f i \gamma^5 f)\phi$ 
    & \textbf{Case (i)} & $a=0$ & ---  \\
\hline
 \multirow{4}{*}{$\chi$ and $V_\mu$} & 
 $\mathcal{L}_5=( g_D \bar \chi \gamma^\mu \gamma^5  \chi + g_f \bar f \gamma^\mu f) V_\mu$ 
 & $a=0$ & \textbf{Case (A)} & Eq.~\eqref{eq:DDL5}  \\

& $\mathcal{L}_6=( g_D \bar \chi \gamma^\mu \gamma^5 \chi + g_f \bar f \gamma^\mu  \gamma^5 f) V_\mu$ & 
\textbf{Case (ii)} & \textbf{Case (A)} & ---   \\

  & $\mathcal{L}_7=(g_D\bar \chi \gamma^\mu  \chi + g_f \bar f \gamma^\mu f) V_\mu$ 
  & \textbf{Case (i)}  & \textbf{Case (C)} & Eq.~\eqref{eq:DDL7}  \\

   & $\mathcal{L}_8=(g_D \bar \chi \gamma^\mu\chi + g_f \bar f \gamma^\mu \gamma^5 f) V_\mu$ 
   & \textbf{Case (i)} & \textbf{Case (C)} & ---  \\
   
\hline
\multirow{4}{*}{$S$ and $\phi$} & $\mathcal{L}_9=(M_{D\phi} S^\dagger S + g_f \bar f f) \phi$ 
& \textbf{Case (i)} & \textbf{Case (B)} & Eq.~\eqref{eq:DDL9}\\

 & $\mathcal{L}_{10}= (M_{D\phi} S^\dagger S + g_f \bar f i \gamma^5 f) \phi$ 
 & \textbf{Case (i)}  & \textbf{Case (B)} & ---  \\
 
 & $\mathcal{L}_{9'}= (g_{D} S^\dagger S \phi + g_f \bar f f) \phi$ 
 & ---  & $b=0$ & ---  \\
 
  & $\mathcal{L}_{10'}= (g_{D} S^\dagger S \phi + g_f \bar f i \gamma^5 f) \phi$ 
 & ---  & $b=0$ & ---  \\
 
\hline

 \multirow{2}{*}{$S$ and $V_\mu$} & $\mathcal{L}_{11}= (i g_D S^\dagger  \overset{\leftrightarrow}{\partial_{\mu}}  S + 
 g_D^2 S^\dagger S V_\mu +
 g_f \bar f \gamma_\mu f) V^\mu$ 
 & $a=0$ & \textbf{Case (C)} & Eq.~\eqref{eq:DDL11}  \\

& $\mathcal{L}_{12}=(i g_D S^\dagger \overset{\leftrightarrow}{\partial_{\mu}} S + 
g_D^2 S^\dagger S V_\mu +
g_f \bar f \gamma_\mu \gamma^5 f) V^\mu $
& $a=0$  & \textbf{Case (C)} & --- \\
\hline
 \multirow{4}{*}{$X_\mu$ and $\phi$} & $\mathcal{L}_{13}=
 (M_{D\phi} X^\mu X_\mu^\dagger + g_f \bar f f) \phi$ 
 & \textbf{Case (i)} & \textbf{Case (D)} & Eq.~\eqref{eq:DDL13}  \\
 
  & $\mathcal{L}_{14}=(M_{D\phi} X^\mu X_\mu^\dagger + g_f \bar f i \gamma^5 f) \phi$ 
  & \textbf{Case (i)} & \textbf{Case (D)} & --- \\
  
  & $\mathcal{L}_{13'}=(g_{D} X^\mu X_\mu^\dagger \phi + g_f \bar f f) \phi$ 
  &---  & $b=0$ & ---  \\
  
  & $\mathcal{L}_{14'}=(g_{D} X^\mu X_\mu^\dagger \phi + g_f \bar f i \gamma^5 f) \phi$ 
  &---  & $b=0$ & ---  \\
  
\hline
 \multirow{4}{*}{$X_\mu$ and $V_\mu$}& $\mathcal{L}_{15}= i g_D\lbrace X^{\mu \nu} X_\mu^\dagger V_\nu  - X^{\mu \nu \dagger } X_\mu V_\nu + X_\mu X^\dagger_\nu V^{\mu \nu}\rbrace $ 
 & \multirow{2}{*}{$a=0$}  & \multirow{2}{*}{\textbf{Case (C)}} & \multirow{2}{*}{Eq.~\eqref{eq:DDL15}}  \\
 & $+ g^2_D\lbrace X^{\dagger}_{\mu}X^{\mu}V_{\nu}V^{\nu} - X^{\dagger}_{\mu}V^{\mu}X_{\nu}V^{\nu} \rbrace + g_f \bar f \gamma^\mu fV_\mu$ & & & \\
 
  & $\mathcal{L}_{16}= i g_D\lbrace X^{\mu \nu} X_\mu^\dagger V_\nu  - X^{\mu \nu \dagger } X_\mu V_\nu + X_\mu X^\dagger_\nu V^{\mu \nu}\rbrace $ 
 & \multirow{2}{*}{$a=0$}  & \multirow{2}{*}{\textbf{Case (C)}} & \multirow{2}{*}{---}  \\
 & $+ g^2_D\lbrace X^{\dagger}_{\mu}X^{\mu}V_{\nu}V^{\nu} - X^{\dagger}_{\mu}V^{\mu}X_{\nu}V^{\nu} \rbrace + g_f \bar f \gamma^\mu \gamma^5 fV_\mu$ & & & \\
\hline \hline
\end{tabular}
\caption{The summary table of all the renormalizable operators for the $Z_2$ even mediator scenario. 
The columns $\sv_{2\mu}$ and $\sv_{4\mu}$ show the cross sections of DM annihilation to 
$2\mu$ and $4\mu$ final states. 
We define $X^{\mu \nu} = \partial^\mu X^\nu - \partial^\nu X^\mu $ and $V^{\mu \nu} = \partial^\mu V^\nu - \partial^\nu V^\mu$.
}
\label{tab:Z2even}
\end{table}

In Table~\ref{tab:Z2even}, all representative interaction types between $Z_2$-even mediator, DM, and $\mu$ are listed. 
Here, we use the notation as defined in Table~\ref{tab:particle_notataion} to present the spin nature of DM and mediator. 
The column $\sv_{2\mu}$ indicates the velocity dependence of the cross section for DM annihilating 
to $\mu^+\mu^-$ final state at the present time. 
However, $\sv_{4\mu}$ is for the process ${\rm DM+DM}\to {\rm MED+MED}$ and then each mediator decays to a pair of muons successively.  
The last column of Table~\ref{tab:Z2even} shows the equation number of the DM-nuclei elastic scattering cross section 
whose formula is given in Appendix~\ref{app:ddformula}. 
We will discuss the DM direct detection in Sec.~\ref{sec:dd}.
The sign "---" in the last column means that   
the cross section is negligible.

Note that some cross sections contain both $s$- and $p$-wave contributions but their $a/b$ ratio is non-trivial. 
Therefore, we divide them into several cases and discuss them below. 
First, for $2\mu$ final state, we can simplify the analytical expressions of $\sigma v$~\cite{1404.0022} near resonance as 
\begin{equation}
    \sigma v \propto \frac{C_0}{(4R-R^2)^2} \left ( \mathcal{C}_1 - \frac{\mathcal{C}_2}{4R-R^2} v^2 \right ),
\end{equation}
where $C_0$ (in $\gev^{-2}$) and $\mathcal{C}_{1,2}$ are positive coefficients. 
The resonance parameter $R$ is defined as $R \equiv (2m_D - M)/ m_D$. 
The conditions $\mathcal{C}_2 v^2 \le \mathcal{C}_1 (4R-R^2)$ is to be kinematically allowed and 
$R\le 2$ is for a physical mass $M$.

\begin{figure}[t]
\includegraphics[width=8cm,height=8.5cm]{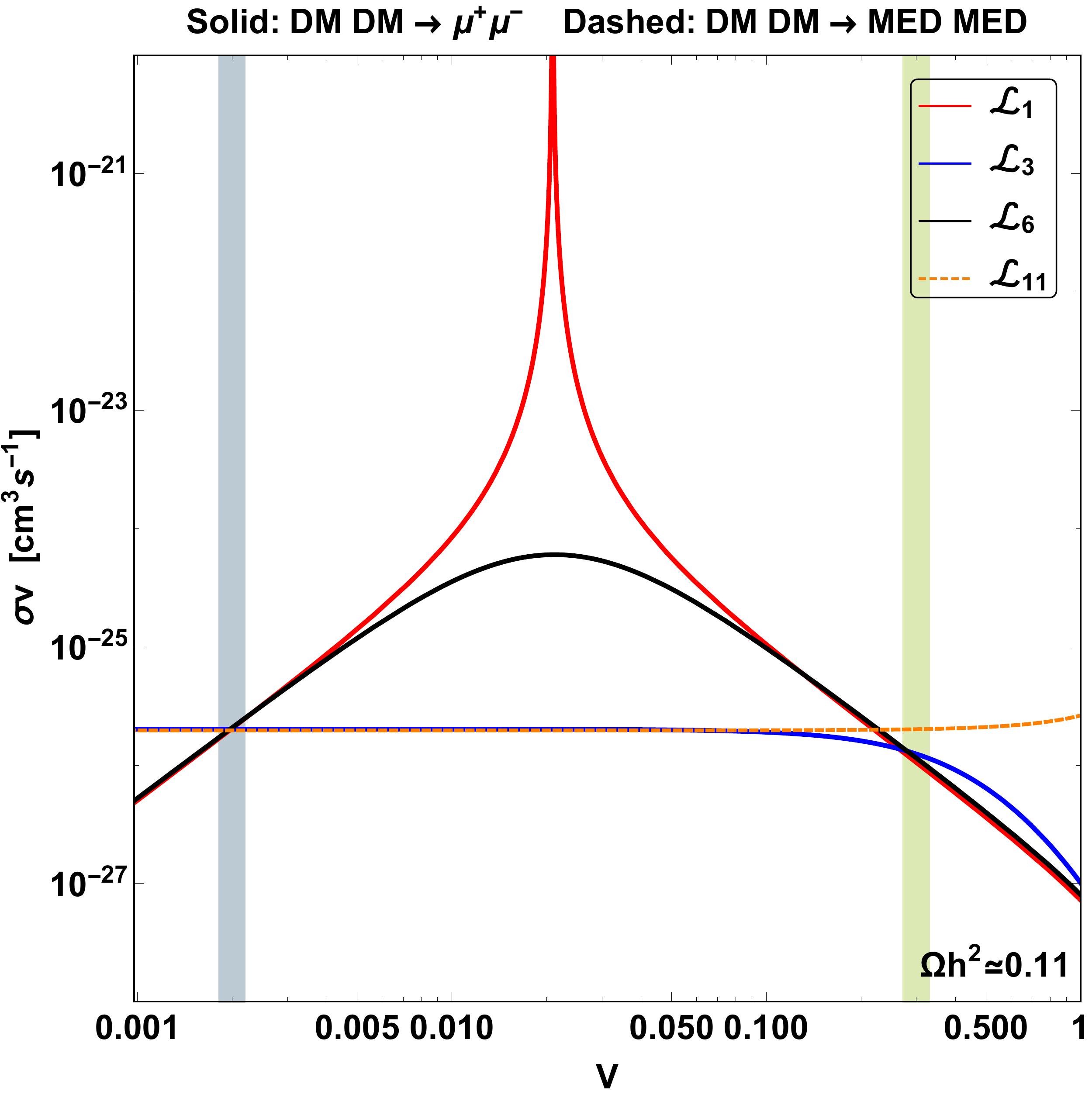}
\includegraphics[width=8cm,height=8.5cm]{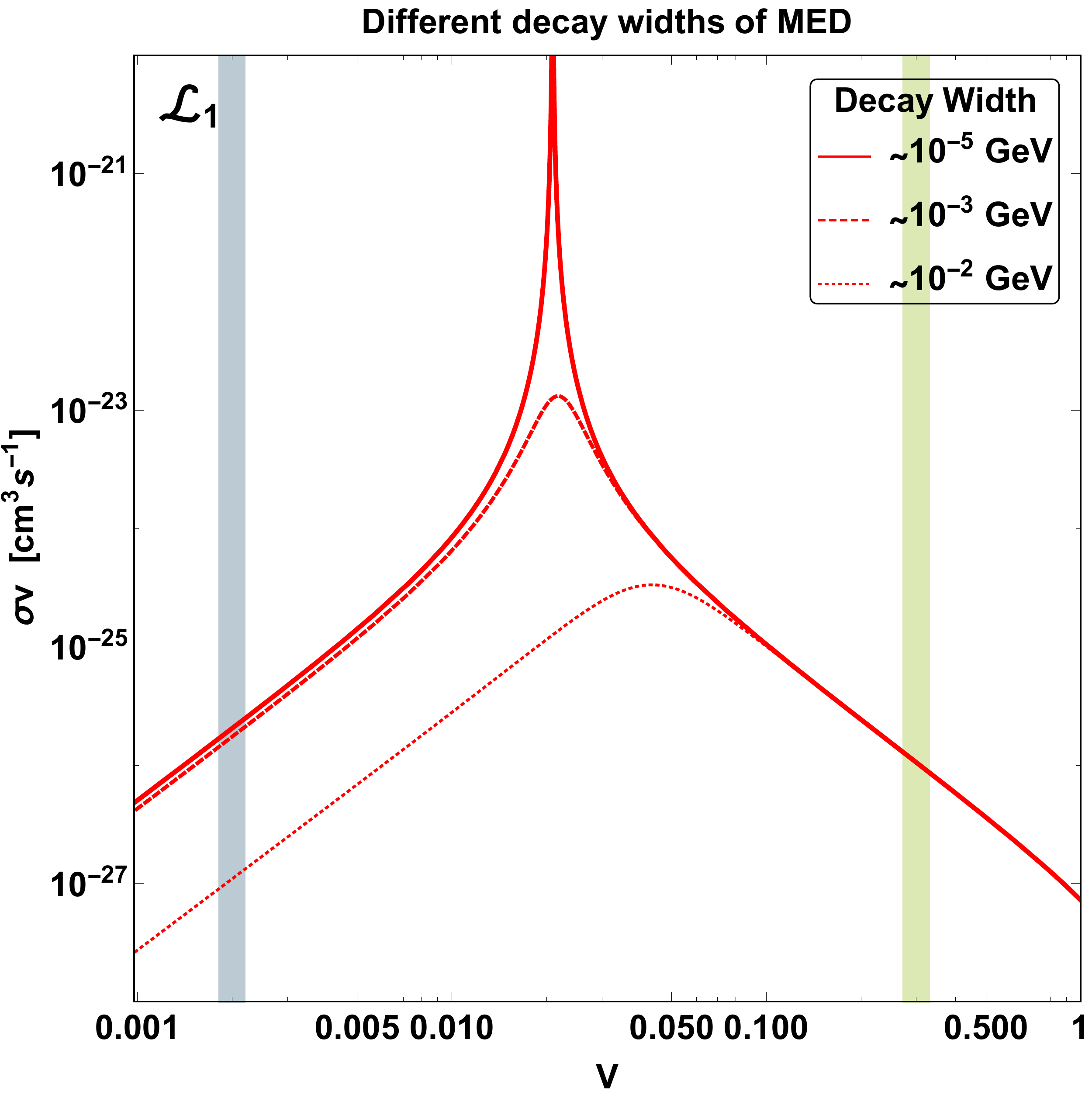}
\caption{The schematic demonstration of $\sigma v$ as function of $v$. 
The parameters of benchmark $\mathcal{L}_{1,3,6}$ are $m_D/\gev=(72.99, 73.11, 73.02)$, $M/\gev=(146.0, 143.5, 146.0)$, 
and $g_D {g_f}\times 10^3=(3.319, 2.494, 3.030)$, respectively. 
For $\mathcal{L}_{11}$, the corresponding parameters are $(m_D,M,g_D)=(68.64~\gev,5.85~\gev,8.53\times10^{-3})$. 
The right panel is the result of $\mathcal{L}_1$ for different decay widths of mediator.
}
\label{fig1:sch_vsv}
\end{figure}

If $0<R\le 2$, we can see $\sigma v$ is with the largest value at $v=0$.  
When $R$ approaches zero from above and the coefficient of $v^2$ is dominant over the first term, 
the total $\sigma v$ with a relativistic speed is smaller than $s$-wave sole component as the blue line 
in the left panel of Fig.~\ref{fig1:sch_vsv}. 
This can explain DM relic density and GCE simultaneously. 
When $0\ll R$, the second term is $v$ suppressed and the cross section is $s$-wave that is  
not able to fit both DM relic density and GCE data. 
On the other hand, the condition $R<0$ implies that annihilating DM needs some kinetic energies to hit the resonance.  
Therefore, we can see the resonance with a small velocity and negative $R$ for $\mathcal{L}_{6}$ and $\mathcal{L}_{11}$
as shown in Fig.~\ref{fig1:sch_vsv}. 
In these negative $R$ regions, one can find a solution for a correct relic density and GCE by 
tweaking the decay width. For $2\mu$ final state, we summarize the correlation between $R$ and $a/b$ 
as follows. 
\begin{itemize}
    \item \textbf{Case (i)}: 
    The sign of $R$ and $a/b$ is always opposite. 
    When we enhance the value of $|R|$, the absolute ratio $|a/b|$ is also increased. 
    The value of $|a|$ and $|b|$ can be comparable around the condition $M/m_D\approx\mathcal{O}(1.3-1.6)$. 
    A smaller $M$ results in $|a/b|>1$ while a larger $M$ yields $|a/b|<1$. 
    \item \textbf{Case (ii)}: Regardless of the sign of $R$, the ratio of $a/b$ is always positive. 
    The correlation between $|R|$ and $|a/b|$ are the same as \textbf{Case (i)}. 
    The condition $|a/b|>1$ only happens at $M\ll m_D$, otherwise $|a/b|<1$ holds.  
\end{itemize}

Unlike $2\mu$ final state, the cross section $\sv_{4\mu}$ behaves in a more sophisticated way. 
To describe the correlation among masses $M, m_D$ and $a/b$, we introduce a different variable $r=M/m_D$ 
and their relations can be summarized into four cases. 
\begin{itemize}
    \item \textbf{Case (A)}: The sign of $a/b$ is positive but there is a maximum value ($a/b\simeq 0.1$) at $r\approx 0.76$. 
    \item \textbf{Case (B)}: In contrast with \textbf{Case (A)}, $a/b$ is always negative while the maximum value ($a/b\simeq -1.5$) happens again at $r\approx 0.76$. The ratio $|a/b|$ is $\mathcal{O}(1)$ but always greater than $1$. 
    \item \textbf{Case (C)}: The ratio $a/b$ is always positive but increased when we reduce the value of $r$. 
    The ratio $a/b\approx\mathcal{O}(1)$ can be found at $r\approx 0.7$.
    
    \item \textbf{Case (D)}: The ratio $a/b$ is always negative. When reducing the value of $r$, $|a/b|$ is enhanced. 
    The ratio $|a/b|\approx\mathcal{O}(1)$ can be found at $r\approx 0.8$. 
\end{itemize}

If DM mass $m_D$ is lighter than the mediator mass $M$, 
the correct relic density can only be achieved by DM annihilation into $2\mu$ final state. 
In the case of $m_D>M$, 
the process DM~$+$~DM~$\to$~MED~$+$MED is kinematically allowed 
so that one has to take both $\sv_{2\mu}$ and $\sv_{4\mu}$ 
contributions into relic density calculation.     
To explain GCE, one has to take a larger annihilation cross section than $\sim 3 \times 10^{-26}~{\rm cm}^3 s^{-1}$ 
which satisfies correct relic density at the early time. 
Thus, the $4\mu$ final state is not favourable to explain both GCE and correct relic density. 
We can see the dashed line as an example in the left panel of Fig.~\ref{fig1:sch_vsv}. 
In addition, 
if $p$-wave $4\mu$ is dominant to DM relic density and $s$-wave $2\mu$ is dominant to GCE, 
it seems to be able to meet the requirements. 
However, the $s$-wave cross section $\sv_{2\mu}\approx 4 \times 10^{-26}~{\rm cm}^3 s^{-1}$ explains GCE data which also implies   
a larger cross section in the early universe than the required value by PLANCK ($\approx 3 \times 10^{-26}~{\rm cm}^3 s^{-1}$).   
Hence, a larger cross section $\sv_{2\mu}\approx 4 \times 10^{-26}~{\rm cm}^3 s^{-1}$ also 
generates the DM relic abundance lower than what we expect.

Note that Majorana and real scalar DM particles are self-conjugate fields, 
thus the DM-DM-Mediator interaction terms have a coefficient of 1/2 and we can simply rescale the value of $g_D$ to compare with Dirac and complex scalar DM scenarios.
However, the vector current does not couple to Majorana or real scalar DM, 
\textit{i.e}. $\mathcal{L}_{7,8}$ and $\mathcal{L}_{11,12}$. 
Therefore, $\mathcal{L}_{7,8}$ and $\mathcal{L}_{11,12}$ will vanish for Majorana and real scalar DM, respectively. 
Besides, $\mathcal{L}_{9^\prime,10^\prime,13^\prime,14^\prime}$ only have DM-DM-MED-MED four point interactions 
which induce only $s$-wave $4\mu$ final states. 
As mentioned previously, these interactions can not explain GCE and DM relic density at the same time.

\begin{figure}[h]
    \centering
    \includegraphics[width=1.0\textwidth]{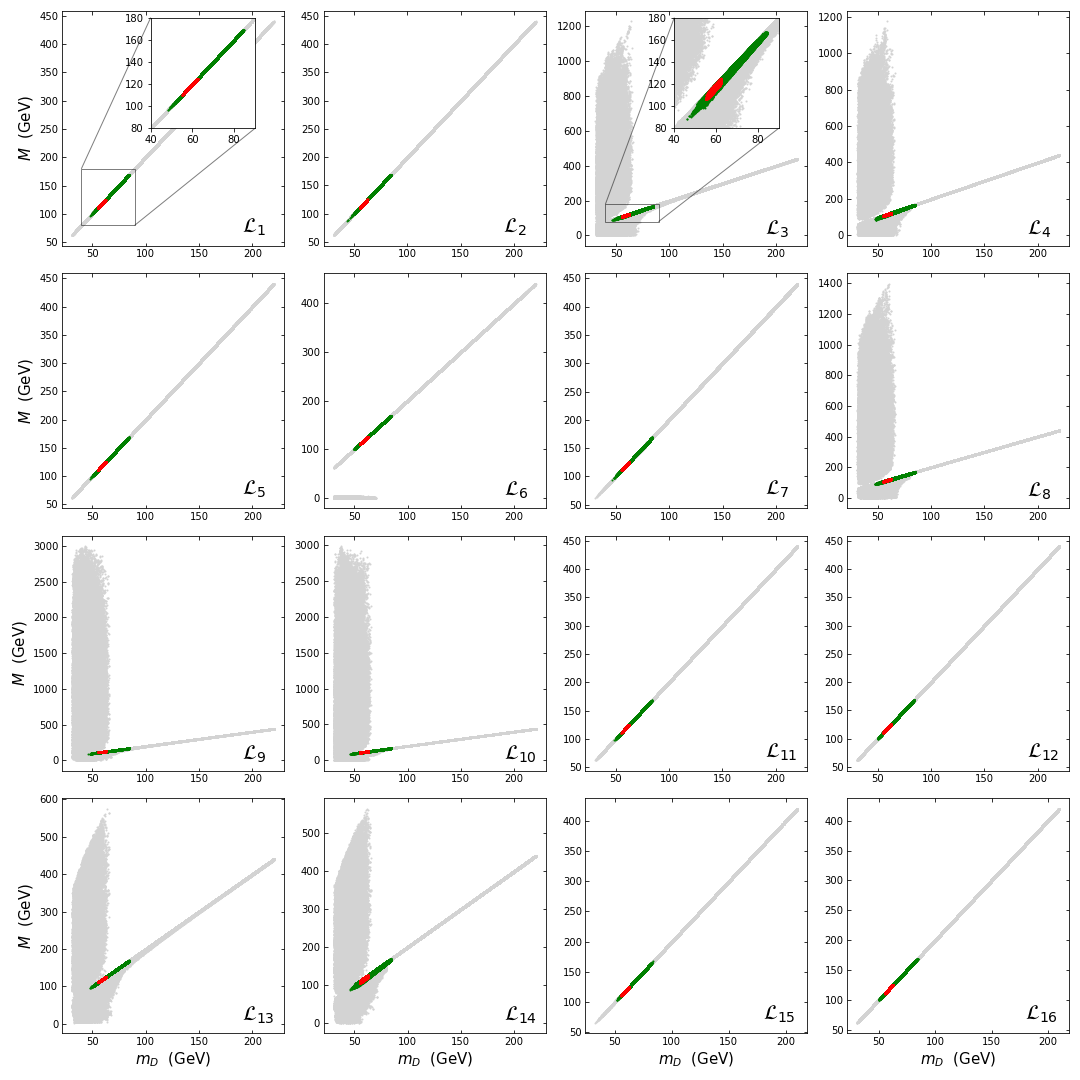}
    \caption{Samples collected with the conditions: $\Delta \chi^2$ within $2\sigma$ confidence level range, projected on ($m_D$, $M$) plane. 
    The red, green, and gray regions correspond to the samples with  
    the original error bars of GCE residual spectra, 
    the error bars amplified by the square root of the minimum reduced chi-square,    
    and the DM spectra fallen into the gray region in Fig.~\ref{fig:sv2sv4}.
    At the lower right corner of each sub-figure, we label the model number and its information is given in Table~\ref{tab:Z2even}.}
    \label{fig:mchivsmphi}
\end{figure}

In the left panel of Fig.~\ref{fig1:sch_vsv}, 
we choose four benchmark interaction types ($\mathcal{L}_1$, $\mathcal{L}_3$, $\mathcal{L}_6$, and $\mathcal{L}_{11}$) 
to present DM annihilation cross section as a function of velocity. 
The light green and blue shaded vertical bands present the peak of   
the DM velocity distributions in the early universe and present Milky Way. 
In the right panel of Fig.~\ref{fig1:sch_vsv}, we qualitatively show three different mediator decay widths: $10^{-5}\gev$ (solid line), 
$10^{-3}\gev$ (dashed line), and $10^{-2}\gev$ (dotted line). 
We find that the peak is so sensitive to the decay width. 
By adjusting the resonance width and height of DM annihilation to $2\mu$,  
it is possible to make the annihilation at the present time higher than the early time.

For each model, we perform several MCMC scans individually to optimize the coverage 
and the parameters are scanned in the following range 
\begin{eqnarray}
     & 20~ {\rm GeV} < m_D < 300~{\rm GeV} , \, 10^{-4}~ {\rm GeV} < M < 3000~ {\rm GeV},  \nonumber \\
     & 10^{-6} < g_f < 2 ,\, 10^{-6} < g_D < 2 , \, 10^{-6}~ {\rm GeV} < M_{D\phi} < 1000~ {\rm GeV}.
\end{eqnarray}

In Fig.~\ref{fig:mchivsmphi}, we show the samples with $\Delta \chi^2 = \chi^2 - \chi^2_{\rm min}$ 
within the confidence level of $2\sigma$ projected to the ($m_D$, $M$) plane. 
The grey samples generate the DM $\gamma$-ray spectra in the grey region of Fig.~\ref{fig:sv2sv4}. 
The $\chi_{\rm GCE}^2$ of these samples in the gray region are taken to be zero, 
otherwise $\chi_{\rm GCE}^2$ is infinity.
The green samples are scaled $\chi_{\rm GCE}^2$ as described in Sec.~\ref{sec:likelihood}, 
while the $\chi_{\rm GCE}^2$ of red samples are calculated via Eq.~\eqref{eq:GCE}.

\begin{figure}[h]
    \centering
    \includegraphics[width=1.0\textwidth]{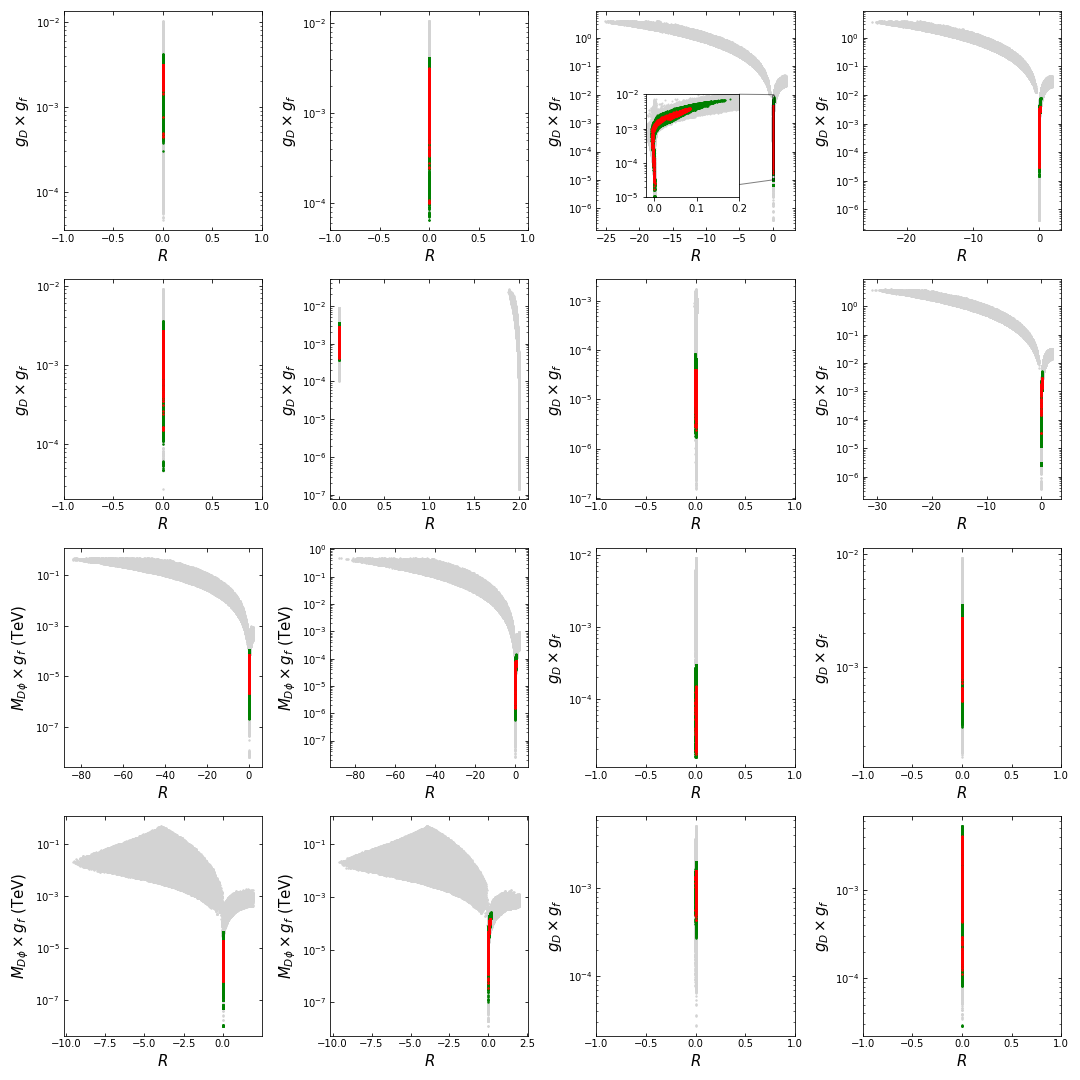}
    \caption{The $2\sigma$ samples on ($R$, $M_{D\phi} g_f$) plane for $\mathcal{L}_{9,10,13,14}$ 
    and ($R$, $g_D g_f$) plane for else.  
    The resonance parameter is defined as $R \equiv (2 m_D - M)/m_D$. 
    The color scheme is same as Fig.~\ref{fig:mchivsmphi}.
}
    \label{fig:mphivsg}
\end{figure}

As shown in Fig.~\ref{fig:mchivsmphi}, $2\sigma$ allowed samples 
for $\mathcal{L}_{1,2,5,6,11,12,15,16}$ locate at the resonance lines, 
because the DM annihilation cross sections are $p$-wave like in these models. 
Only the mediator resonance region can fulfill both DM relic density and GCE constraints. 
Although the samples of $\mathcal{L}_{7}$ show the same characteristic, 
DM in this model cannot be $p$-wave annihilation. 
For $\mathcal{L}_{6}$, there is a tiny fraction of samples survived 
at the region ($M\ll m_D$). 
As described in \textbf{Case (ii)}, annihilation cross section behaves as $s$-wave when $M \ll m_D $. 
The rest of the models are $s$-wave and there are large allowed areas at $m_D \sim 30-80$ GeV. 
The DM masses in all models are extended to as high as $m_D \sim 230\gev$. 
As the DM mass is increased, we need a larger $\sv$ 
to enhance the gamma-ray flux ($Q \sim\sv/m_D^2$) to avoid the lower bound of 
the GCE fluxes.  
However, such a larger $\sv$ in the present universe can be reached only via resonance.

Apart from the highly resonant regions, 
in some models, \textit{i.e.} $\mathcal{L}_{3,4,6,8,9,10,13,14}$, 
broader allowed parameter spaces are exist near below the $M = 2 m_D$ line. 
We only zoomed in the $\mathcal{L}_{1}$ and $\mathcal{L}_3$ panels 
of Fig.~\ref{fig:mchivsmphi} for a demonstration.  
To explain this more clearly, 
we also present Fig.~\ref{fig:mphivsg} as complementary to Fig.~\ref{fig:mchivsmphi}
which shows the relevant couplings in a function of $R \equiv (2 m_D - M)/m_D$. 
We can see the resonant annihilation region located at the narrow vertical strip regions
where DM annihilation to $2\mu$ behaves as $p$-wave in Table \ref{tab:Z2even}.
Therefore, to explain both signals (GCE and relic density), 
the very fine-tuning parameter space, namely resonance region, is needed, 
regardless of $s$, $p$ or $s+p$ wave, 
unless the GCE systematic uncertainties have been taken into account.

\subsection{$Z_2$-odd mediator}
\label{sec:z2odd}
\begin{table}[t]
    \centering
    \begin{tabular}{c|c|c|c|c}
\hline\hline
\multicolumn{5}{c}{$Z_2$ odd mediator} \\
\hline \hline
 types  & Lagrangian & $\langle \sigma v\rangle_{2\mu}$ & DM field & DD  \\
\hline \hline
$\chi$ and $\phi$  & $\mathcal{L}_{17}=g_D \bar \chi P_R f \phi +$h.c.
& $s$ & Dirac & Eq.~\eqref{eq:DDL17}  \\

\hline
$\chi$ and $V_\mu$ & $\mathcal{L}_{18}=g_D \bar \chi \gamma^\mu P_R f V_\mu +$h.c. 
    & $s$ & Dirac & Eq.~\eqref{eq:DDL18}  \\

\hline
$\chi$ and $\phi$ & 
 $\mathcal{L}_{19}=g_D \bar \chi P_R f \phi+$h.c. 
 & $p$ & Majorana & suppressed  \\

\hline
$\chi$ and $V_\mu$
  & $\mathcal{L}_{20}=g_D\bar \chi \gamma^\mu  P_R f V_\mu+$h.c. 
  & $p$  & Majorana & suppressed \\
   
\hline
$S$ and $\psi$ & $\mathcal{L}_{21}=g_D \bar \psi P_R f S +$h.c. 
& \textbf{Case (i)} & Real & suppressed  \\

\hline

$S$ and $\psi$ & $\mathcal{L}_{22}= g_D \bar \psi P_R f S+$h.c. 
 & $p$ & Complex & Eq.~\eqref{eq:DDL22}  \\

\hline
$X_\mu$ and $\psi$
 & $\mathcal{L}_{23}=g_D \bar \psi \gamma^\mu P_R f X_\mu^\dagger+$h.c. 
 & $s$ & Real/Complex & suppressed  \\
 
\hline \hline
\end{tabular}
\caption{All representative interaction types among $Z_2$-odd mediator, DM and right handed muon. 
The operator $P_R = (1 + \gamma^5)/2$ refers to the right-handed projection.
}
\label{tab:Z2odd}
\end{table}

For $Z_2$-odd mediator, it must carry the electric charge in muonphilic DM models.  
Hence, we require that the mass of the mediator has to be heavier than DM 
to prevent DM from decaying to the charged mediator. 
In Table~\ref{tab:Z2odd}, we list all representative renormalizable interaction types with the right handed muon. 
Notice that we only consider the SM singlet mediator fields in this study as the minimal models. 
If the mediator fields are SM doublets or even more complicated extensions with new gauge symmetries, 
they can couple to left handed muon or both 
chiralities 
as well. 
Because the mediator mass is always heavier than DM, 
the only possible annihilation final state is $2\mu$. 
Compared with the $Z_2$-even mediator case, there is 
no resonance enhancement. 
Thus, we are safe to exclude the $p$-wave interactions $\mathcal{L}_{19,20,22}$ because they are not able to simultaneously generate  
the correct relic density and the DM annihilation cross section required by GCE.

The charged mediator such as slepton suffers from 
the stringent lower mass limit $103.5\gev$ from LEP~\cite{ALEPH:2002gap}.  
When the charged mediator mass is heavier than LEP limit, 
it implies that the larger coupling $g_D$ is needed to fit 
the DM relic density and GCE cross section. 
Consequently, such a considerable coupling may violate 
the PandaX 4T and XENON1T limit.  
For the $Z_2$-odd mediator scenario, we perform the parameter scans in the following range 
\begin{eqnarray}
     & 20~ {\rm GeV} < m_D < 200~{\rm GeV} , \, m_D < M < 1000~ {\rm GeV}, \, 10^{-6} < g_D < 2. \nonumber 
\end{eqnarray}

\begin{figure}[h]
\centering
\includegraphics[width=0.65\textwidth]{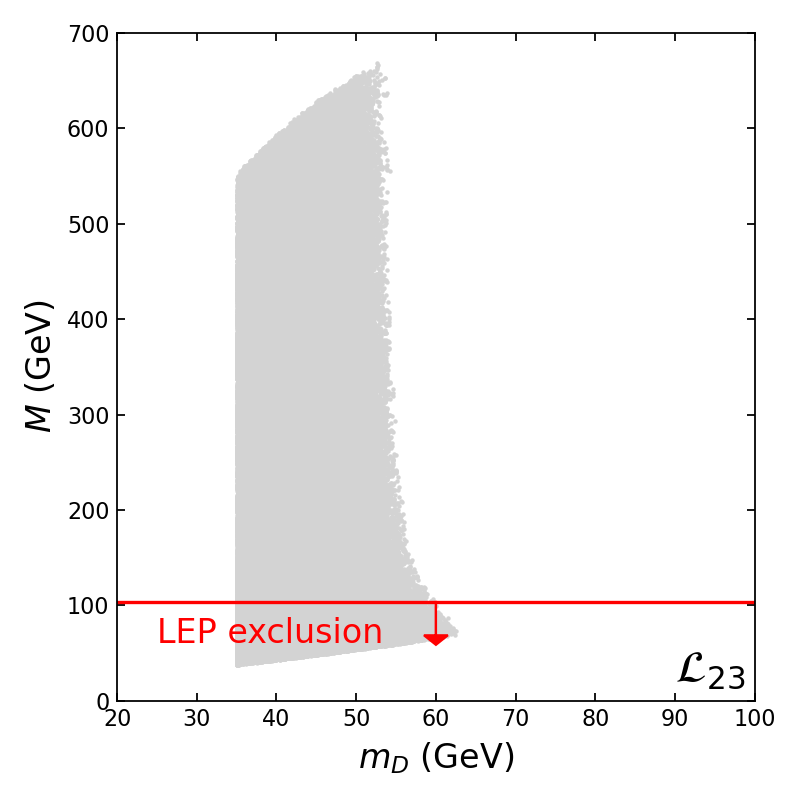}
\caption{Samples with $\Delta \chi^2$ in 2$\sigma$ region for $\mathcal{L}_{23}$. 
The horizontal red solid lines are the LEP upper limit 103.5 GeV~\cite{ALEPH:2002gap}. 
Only the gray region where the conservative systematic uncertainties are taken into account  
can escape from the LEP limit.}
\label{fig:tch}
\end{figure}

In Fig.~\ref{fig:tch}, we show our results for $\mathcal{L}_{23}$ on ($m_D$, $M$) plane. 
The $s$-wave type interactions $\mathcal{L}_{17,18}$ are incompatible with current DD data.
On the other hand, $\sv$ of $\mathcal{L}_{21}$ behaves like the one of $\mathcal{L}_{3}$. 
Thus, there can be some samples are survived from all the constraints except for 
the LEP mass limit. 
The LEP experiment has set a severe limit (the vertical red solid line) 
to exclude the light charged particles. 
Hence, most of the $Z_2$-odd mediator scenarios cannot explain GCE and DM relic density 
at the same time but $\mathcal{L}_{23}$ has a viable parameter space only when we take 
the conservative systematic uncertainties into account.

\section{Discussion of DM direct detection}
\label{sec:dd}

\begin{figure}[h]
\centering
\includegraphics[width=0.8\textwidth]{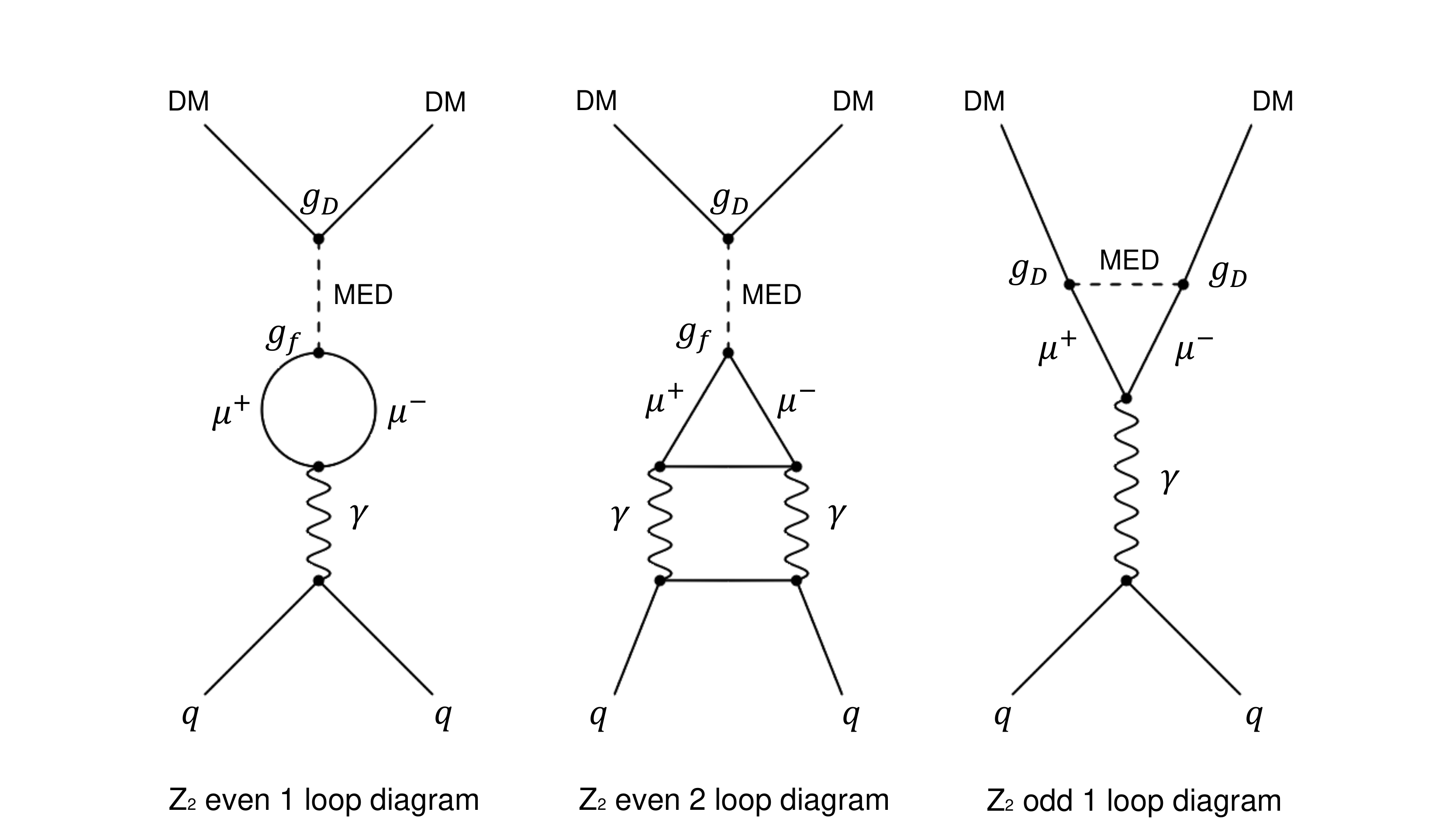}
    \caption{The Feynman diagram for the loop-induced DM-quark scattering. 
    The one loop contribution for the $Z_2$-even mediator scenario is presented in the left panel while 
    the two loop contribution is given in the middle one. 
    For the $Z_2$-odd mediator scenario, we draw the one loop contribution in the right panel.}
\label{fig:feynman}
\end{figure}

For the simplified muonphilic DM models, there is no tree level DM-nuclei elastic scattering. 
The relevant Feynman diagrams of the loop-induced DM-quark scattering are depicted in Fig.~\ref{fig:feynman}. 
First, we consider the $Z_2$-even mediator case and  
define the general lepton current as $\overline{l}\Gamma_l l$. 
Following Ref.~\cite{Agrawal:2014ufa}, 
the one loop contributions are nonzero only for vector and tensor lepton currents, namely $\Gamma_l = \gamma_{\mu}, \sigma_{\mu\nu}$. 
Therefore, only $\mathcal{L}_{5,7,11,15}$ can generate one loop contributions to 
the DM-nuclei elastic scattering~\cite{Kopp:2009et,Duan:2017pkq,Athron:2017drj,YaserAyazi:2019psw}  
as shown in the left panel of Fig.~\ref{fig:feynman}.

For the scalar lepton current, $\Gamma_l = 1$, the one loop contribution vanishes since a scalar current cannot couple to a vector current. 
The DM-quark interaction can only be induced at two loop level for $\mathcal{L}_{1,3,9,13}$~\cite{Athron:2017drj,YaserAyazi:2019psw}, 
as depicted in the middle panel of Fig.~\ref{fig:feynman}.

For pseudo-scalar and axial-vector lepton currents $\Gamma_l = \gamma_5, \gamma_{\mu}\gamma_5$, 
the diagrams vanish to all loop orders. 
The interaction with $\gamma_5$ gives either zero or a fully anti-symmetric tensor $\epsilon^{\alpha\beta\mu\nu}$. 
Since there are only three independent momenta in the $2\rightarrow 2$ scattering process, 
two indices can be contracted with the same momentum and a zero amplitude square is obtained. 
Therefore, we mark ``---'' in the last column of Table.~\ref{tab:Z2even} for $\mathcal{L}_{2,4,6,8,10,12,14,16}$.

For the case of $Z_2$-odd mediator, the DM-nuclei scattering cross sections are suppressed 
for the self-conjugate DM, namely real scalar, Majorana fermion, and real vector. 
As given in Ref.~\cite{Agrawal:2014ufa}, the self-conjugate DM couples to a single photon in $t$-channel simplified models 
only through the anapole moment. 
This leads to that DM-quark scattering amplitude is suppressed in the non-relativistic limit as for $\mathcal{L}_{19,20,21,23}$.  
On the other hand, if the muonphilic DM are complex scalar, Dirac fermion and complex vector, 
the one loop induced DM-quark interactions cannot be ignored~\cite{Agrawal:2014ufa,Agrawal:2011ze,Bai:2014osa},
and the Feynman diagram shown in the right panel of Fig.~\ref{fig:feynman}.

\begin{figure}
    \centering
    \includegraphics[width=0.85\textwidth]{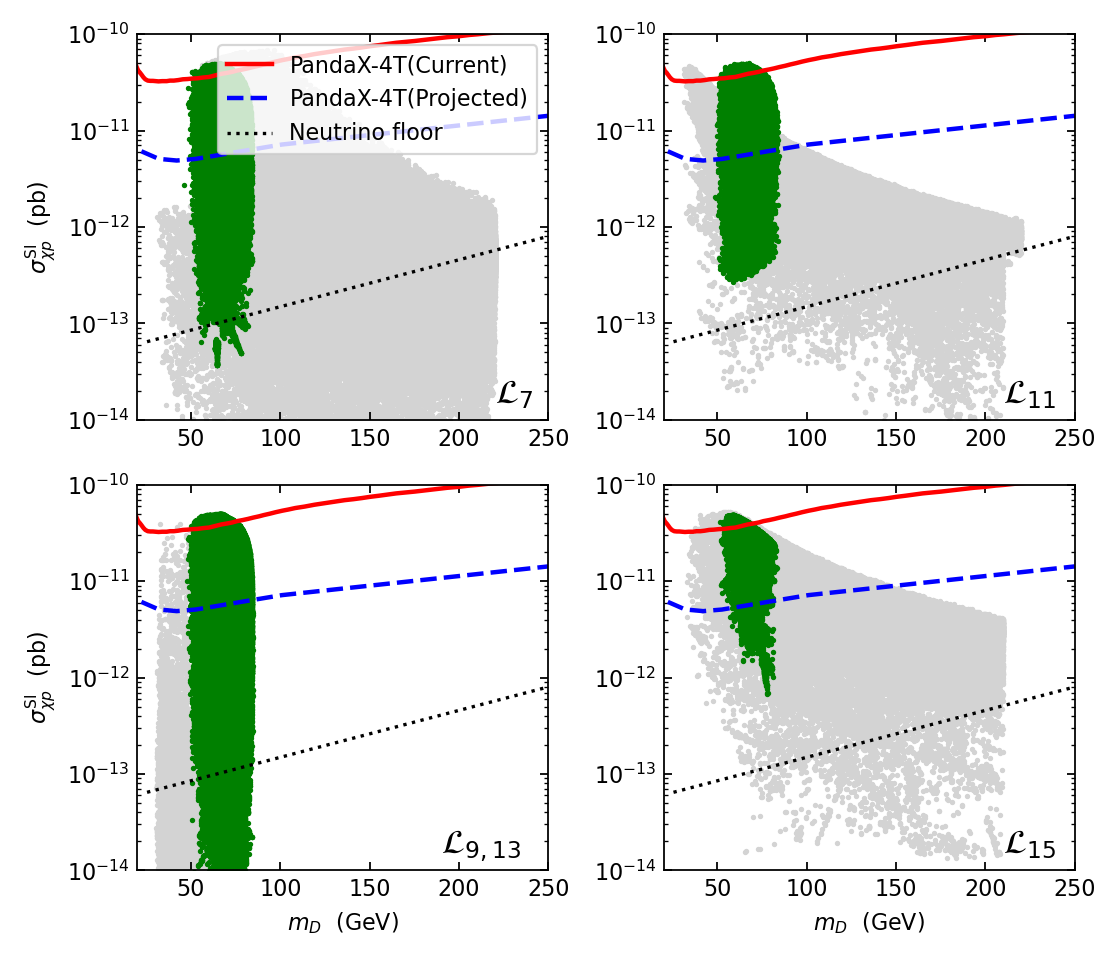}
    \caption{The $2\sigma$ distribution on the ($m_D$, $\sigsip$) plane.
    The red solid lines are current PandaX-4T $90\%$ upper limit~\cite{2107.13438}, 
    the blue dashed lines are expected PandaX-4T $90\%$ upper limit~\cite{1806.02229}, 
    and the black dotted lines are neutrino floor~\cite{Vergados:2008jp}. 
    The color coding is same as Fig.~\ref{fig:mchivsmphi}.
    }
    \label{fig:SI}
\end{figure}

In Fig.~\ref{fig:SI}, we plot $2\sigma$ distributions on the ($m_D$, $\sigsip$) plane 
for five models $\mathcal{L}_{7,9,11,13,15}$ where $\mathcal{L}_{9}$ and $\mathcal{L}_{13}$ 
are fairly similar to each other.  
The red thick lines are the $90\%$ upper limits of present PandaX-4T while its future sensitivity~\cite{1806.02229} is presented by 
blue thick dashed lines. The neutrino floor is demonstrated by the black dotted lines.  
Except for $\mathcal{L}_{7,9,11,13,15}$, other models are not shown here because their cross sections are below the neutrino floor.  
The common feature of $\mathcal{L}_{9}$ and $\mathcal{L}_{13}$ is with dimensional coupling $M_{D\phi}$ 
which makes an optimistic prediction of $\sigsip$ in spite of two loop level contributions. 
On the other hand, $\sigsip$ for $\mathcal{L}_{7}$, $\mathcal{L}_{11}$ and $\mathcal{L}_{15}$ 
are only one loop level and not further suppressed by the fine-structure constant.

\section{The muon $g-2$ excess}
\label{sec:g2}

\begin{figure}
    \centering
    \includegraphics[width=0.8\textwidth]{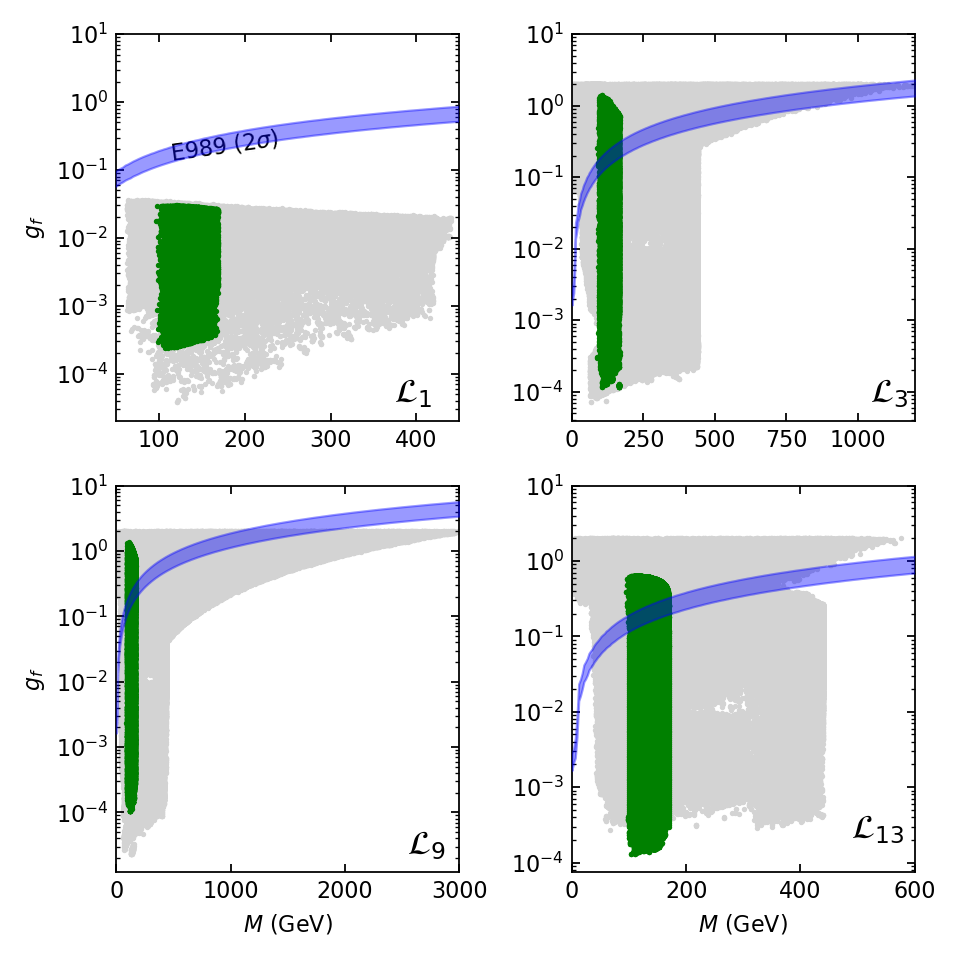}
    \caption{The $2\sigma$ distributions in the ($M$, $g_f$) plane. 
    The same color coding as Fig.~\ref{fig:mchivsmphi} is used.
    The purple shaded belt corresponds to the E989 $2\sigma$ region of $\delta a_\mu$ in each model.}
    \label{fig:gm2}
\end{figure}

The muon anomalous magnetic moment $a_\mu$ has been recently announced by E989 at Fermilab. 
By combining the new data with the previous measurement from Brookhaven National Lab (BNL)~\cite{ParticleDataGroup:2018ovx}, 
they found a deviation $\delta a_\mu = (2.51 \pm 0.59) \times 10^{-9}$ with 4.2$\sigma$ significance~\cite{Muong-2:2021ojo}
 from the value of the SM prediction.

Since the mediator couples to muon lepton, we intuitively check whether  
the mediator in the loop can contribute to the experimentally measured muon $g-2$ excess. 
In the Appendix~\ref{append:gm2}, we provide four analytical formulas for scalar, pseudo-scalar, vector, and axial-vector mediators. 
The contributions from pseudo-scalar and axial-vector mediator are negative at one loop level.  
For vector mediator, $\delta a_\mu$ is too small to reach $2\sigma$ region. 
Thus, only the contributions form scalar mediators are shown in Fig.~\ref{fig:gm2}. 
When muon-muon-mediator coupling $g_f$ is around the order of $\sim 0.1$, muon $g-2$ excess can be explained. 
\textit{Therefore, as long as the E989 result can be confirmed in the near future, only $\mathcal{L}_{3}$ (fermionic DM), $\mathcal{L}_{9}$ (scalar DM) and 
$\mathcal{L}_{13}$ (vector DM) are allowed to explain the correct DM relic density, GCE and muon $g-2$ excess simultaneously. }

\section{Conclusions}
\label{sec:conclusions}

The GCE is a well-known anomaly and the DM annihilation explanation is one of the popular solutions. 
If this DM annihilation explanation is correct, the same DM mass and cross section shall be found 
in other astrophysical observations such as Fermi dSphs $\gamma$ data and AMS-02 $e^+$ and $\bar{p}$ 
data. Especially, Ref.~\cite{DiMauro:2021qcf} has suggested that $\chi\chi\to\mu\mu$ annihilation 
can explain all the astrophysical observations consistently. 
Motivated by such a claim, we perform a comprehensive analysis for the muonphilic DM. 
We attempt to probe the muonphilic DM from the particle physics point of view. 
We first build 16 interaction types for $Z_2$-even mediator ($s$-channel and contact interaction annihilation) 
while 7 interaction types for $Z_2$-odd mediator ($t$-channel annihilation). 
We hire the measurement from Fermi GCE, PLANCK relic density, PandaX-4T $\sigsip$, and LEP new charged particle search 
in order to pin down the favoured interaction types and their parameter space.  
After we find the allowed interaction types and parameter space, we discuss their predictions of muon $g-2$ by comparing 
with the recently reported excess from the FermiLab E989 experiment.

In this work, we demonstrate our results by using two types of GCE systematic uncertainties. 
The first type (``normal approach") is to simply enlarge the error bars given in~\cite{DiMauro:2021qcf} 
with the square-root of the minimum reduced chi-square. 
This trick enables us to find out the potentially overlooked systematic uncertainties. 
The normal approach is presented by green color. 
However, we also presented the results with the second type GCE systematic uncertainties as   
a more conservative error band (gray band) shown in Fig.~\ref{fig:sv2sv4}.  
The results using the second type uncertainties are called ``conservative approach" 
and presented by gray color.

We summarize our findings for muonphilic DM models with $Z_2$-even mediators.
First, we find that all of 16 interaction types with $Z_2$-even mediators, 
apart from four-point interactions, can provide correct DM relic density, 
an explanation to GCE but a tiny $\sigsip$ to DM DD. 
For normal approach, only the narrow phase spaces of resonances are remained 
to accommodate both GCE and DM relic density. 
Because the favoured $\sv$ for GCE is higher than the one from the relic density, 
both the velocity-independent cross section ($s$-wave) and the $v^2$-dependent cross section ($p$-wave) 
cannot be the solutions. 
Additionally, the annihilation to $4\mu$ final state requires 
a cross section $\sv_{4\mu}\sim 8\times 10^{-26}$~cm$^3 s^{-1}$ to explain GCE. 
Such a cross section is higher than the one required by the correct DM relic density. 
Therefore, if considering normal approach, 
we are unable to find any solution for $M<m_D$ where DM annihilate to a pair of mediators. 
On the other hand, for $M>m_D$, we still find some solutions with two mechanisms in a subtle way. 
The first mechanism is resonance based on the condition that the mediator is lighter than twice DM mass. 
By tuning the decay width, one can simply tweak the resonance position to correctly obtain 
the early and present DM cross sections. 
The second mechanism happens only for the condition $2 m_D> M$. 
Under this condition, one can find some cancellation between $s$-wave and $p$-wave 
so that $\sv$ at the early universe time can be suppressed.  
Clearly, if the annihilation cross section only contains the $p$-wave contribution, 
the allowed parameter space only owes to the first resonance mechanism. 
However, for conservative approach, the constraints are less tight. 
A sizable amount of non-resonant samples appears around $m_D \sim 60$ GeV 
in interactions $\mathcal{L}_{3,4,8,9,10,13,14}$ and the DM masses can be as higher as 200 GeV.

As for the muonphilic DM models with $Z_2$-odd mediators,
because a $Z_2$-odd mediator must carry an electric charge in these models, the condition $m_D<M$ is held to maintain the electrically neutral universe. 
To explain the DM annihilation cross sections required by the GCE and PLANCK relic density measurement, 
it results a lighter mediator and a larger coupling. 
However, the PandaX-4T $\sigsip$ upper limit already rules out most of the interaction types 
due to their larger couplings required to explain GCE.  
Only the interaction $\mathcal{L}_{21}$ and $\mathcal{L}_{23}$ are survival 
from the PandaX-4T constraints but the required mediator masses of $\mathcal{L}_{21}$ 
are still lower than the LEP new charge particle mass limit. 
Therefore, all of interaction types with $Z_2$-odd mediators are excluded based on normal approach. 
For conservative approach, $\mathcal{L}_{23}$ is still survived.

Although the muonphlic DM can only scatter with proton via loop contributions, 
the current PandaX-4T $\sigsip$ upper limit is still sensitive to them. 
If considering those dimensional couplings $M_{D\phi}$, even the two loop contribution can be largely probed before 
reaching the neutrino floor. 
Besides $\mathcal{L}_{7,9,11,13,15}$, the rest interaction types are still hidden below the neutrino floor. 
Similarly, if muon $g-2$ result from E989 can be confirmed, 
only the scalar mediator is allowed and the possible interaction types are 
$\mathcal{L}_{3}$ (fermionic DM), $\mathcal{L}_{9}$ (scalar DM) and $\mathcal{L}_{13}$ (vector DM). 
Among these three models, only $\mathcal{L}_{3}$ cannot be tested by future DD experiments. 

Finally, we would like to comment the GeV anti-proton excess in the AMS-02 data 
and searches of the allowed muonphilic DM models at the future muon collider.  
In this work we solely follow the argument of Ref.~\cite{DiMauro:2021qcf} 
on the absence of the anti-proton excess. 
The anti-proton created by muon DM annihilation final state can be neglected. 
If there was a GeV anti-proton excess as argued in \cite{2017PhRvL.118s1101C,2017PhRvL.118s1102C}, 
the GCE as well as the $g-2$ anomaly would also shed valuable light on the possible DM origin~\cite{Abdughani:2021pdc}. 
On the other hand, the proposed 3 TeV muon collider~\cite{MuonCollider:2022xlm} is an ideal machine to test the MED and DM for the survival parameter space in this study. The $Z_2$-even ($Z_2$-odd) MED can be searched for via $\mu^+ \mu^-\rightarrow\mu^+ \mu^-$ (MED pair production) process. The mono-$\gamma$ process is used to explore DM for both $Z_2$-even and $Z_2$-odd MED models. 
We will return to a detailed study of the muon collider in a future work. 


\section*{Acknowledgments}

This work was supported by the National Natural Science Foundation of China (U1738210, 11921003, 12047560), China Post-doctoral Science Foundation (2020M681757), Chinese Academy of Sciences.

\appendix

\section{Inverse Compton scattering from DM annihilation}
\label{sec:ICS}

Generally, the electron/positron $e^\pm$ spectrum for ICS is obtained by solving the diffusion equation
\begin{equation}
    \frac{\partial}{\partial t} \frac{\partial n_e}{\partial E} = 
    \nabla \left[ D(E,\mathbf{r}) \nabla \frac{\partial n_e}{\partial E} \right] + 
    \frac{\partial}{\partial E} \left[ b(E,\mathbf{r}) \frac{\partial n_e}{\partial E} \right] + 
    Q(E,\mathbf{r}),
\end{equation}
where $\frac{\partial n_e}{\partial E}$ is the equilibrium electron density in the interval $d^3\mathbf{r}dE$. 
The diffusion coefficient and the energy loss rate are defined as $D(E,\mathbf{r})$ and $b(E,\mathbf{r})$, respectively. 
The electron source term, $Q(E,\mathbf{r})$, can be expressed as
\begin{equation}
    Q(E,r) = \frac{\langle \sigma v \rangle \rho^2(r)}{2m_D^2} \frac{dN_e}{dE_e}.
\end{equation}
Here $\langle \sigma v \rangle$ is DM averaged annihilation cross section, $\rho(r)$ is DM density profile, $m_D$ is DM mass, and $\frac{dN_e}{dE_e}$ is $e^\pm$ injection spectrum per DM annihilation. If DM is not self-conjugated, $\langle \sigma v \rangle$ will be replaced by $\langle \sigma v \rangle$/2.

The local emissivity of $\gamma$ rays is defined as
\begin{equation}
    j_{\rm ICS}(E_\gamma,r) = 2\int_{m_e}^{M_D} dE \frac{dn_e}{dE} P_{ICS}(E,E_\gamma),
\end{equation}
and the ICS power $P_{ICS}(E,E_\gamma)$ is given by
\begin{equation}
    P_{ICS}(E,E_\gamma) = E_\gamma \int d\epsilon n(\epsilon) \sigma (E_\gamma,\epsilon,E),
\end{equation}
where $\epsilon$ is the energy of target starlight, $n(\epsilon)$ is photon number density. $E$ and $E_\gamma$ are the energy of electrons/positions and upscattered photons, respectively. The ICS cross section $\sigma (E_\gamma,\epsilon,E)$ can be written as
\begin{equation}
    \sigma (E_\gamma,\epsilon,E) = \frac{3\sigma_T}{4\epsilon \gamma^2} G(q,\lambda),
\end{equation}
which is the so-called Klein-Nishina formula. Here $\sigma_T \sim 0.665$ barn is the Thomson cross section, and $G(q,\lambda)$ is given by~\cite{Blumenthal:1970gc}
\begin{equation}
    G(q,\lambda) = 2q\ln q + (1+2q)(1-q) + \frac{(2q)^2(1-q)}{2(1+\Gamma q)},
\end{equation}
where
\begin{equation}
    \Gamma = \frac{4\gamma^2 \epsilon}{E}, ~~ q=\frac{E_\gamma}{\Gamma (E-E_\gamma)}.
\end{equation}


Finally, the approximation of integrated flux density for ICS at energy $E_\gamma$ for a small region with much greater distance than size is
\begin{equation}
    \frac{d\Phi_\gamma}{dE_\gamma} \approx \frac{1}{D_A^2} \int dr r^2 \frac{j_{\rm ICS} (E_\gamma,r)}{E_\gamma},
\end{equation}
in which $D_A$ is the angular diameter distance.

In this work, we adopted simplified power-law diffusion coefficient
\begin{equation}
    D(E) = D_0 E^\gamma
    \label{eq:diffusion}
\end{equation}
with diffusion constant $D_0 = 3 \times 10^{28}$ cm$^2$s$^{-1}$ and $\gamma = 0.3$. The photon number density $n(\epsilon)$ 
for starlight is taken as black body model
\begin{equation}
    n(\epsilon) \propto \frac{8\pi \epsilon^2}{e^{\epsilon/T}-1}
\end{equation}
with the temperature $T = 3500$ K.

For the DM prompt $\gamma$ emission, the flux can be calculated as  
\begin{equation}
    \frac{d\Phi_\gamma}{dE_\gamma}=\frac{1}{4\pi}\frac{\sv}{2 m_D^2} \dndEga \times \int_{\Delta\Omega} \int_{\rm l.o.s.}dld\Omega \rho^2(\mathbf{r}(l,\mathbf{n})), 
    \label{eq:promt}
\end{equation}
in which the integration is taken along the line-of-sight (l.o.s.) toward the GC.

\section{DM direct detection cross sections}
\label{app:ddformula} 

Here we show all non-zero DM-nucleus SI cross sections for the simplified muonphilic DM models in Table \ref{tab:Z2even} and \ref{tab:Z2odd}~\cite{Kopp:2009et,Agrawal:2014ufa,Duan:2017pkq,Athron:2017drj,YaserAyazi:2019psw} :

\begin{eqnarray} 
\sigma^\mathrm{SI}_{\mathcal{L}_1} &=& \sigma_0 \left(\frac{\pi \alpha Z \mu_N v}{6\sqrt{2}} \right)^2 
\left(\frac{g_D g_f}{m_\mu} \right)^2,  \label{eq:DDL1} \\
\sigma^\mathrm{SI}_{\mathcal{L}_3} &=& \sigma_0 \left(\frac{\pi \alpha Z \mu_N v}{6\sqrt{3}} \right)^2 
\left(\frac{g_D g_f \mu_N v}{m_\mu m_D} \right)^2,  \label{eq:DDL3} \\ 
\sigma^\mathrm{SI}_{\mathcal{L}_5} &=& \sigma_0 \left( \frac{g_Di g_f}{3} \right)^2 \left( \ln \frac{m_\mu^2}{\Lambda^2} \right)^2 v^2\left(1+\frac{\mu^2_N}{2 m^2_N}\right), 
\label{eq:DDL5} \\
\sigma^\mathrm{SI}_{\mathcal{L}_7} &=& \sigma_0 \left( \frac{g_D g_f}{3} \right)^2 \left( \ln \frac{m_\mu^2}{\Lambda^2} \right)^2, \label{eq:DDL7} \\
\sigma^\mathrm{SI}_{\mathcal{L}_9} &=& \sigma_0 \left( \frac{\pi \alpha Z \mu_N v}{12\sqrt{2}} \right)^2 \left( \frac{M_{D\phi} g_f}{m_\mu m_D} \right)^2, 
\label{eq:DDL9} \\
\sigma^\mathrm{SI}_{\mathcal{L}_{11}} &=& \sigma_0 \left( \frac{g_D g_f}{24} \right)^2 \left( \ln \frac{m_\mu^2}{\Lambda^2} \right)^2, \label{eq:DDL11} \\
\sigma^\mathrm{SI}_{\mathcal{L}_{13}} &=& \sigma_0 \left( \frac{\pi \alpha Z \mu_N v}{6\sqrt{2}} \right)^2 \left( \frac{2 M_{D\phi} g_f}{m_\mu m_D} \right)^2,  \label{eq:DDL13} \\ 
\sigma^\mathrm{SI}_{\mathcal{L}_{15}} &=& \sigma_0 \left( \frac{g_D g_f}{8} \right)^2 \left( \ln \frac{m_\mu^2}{\Lambda^2} \right)^2, \label{eq:DDL15} \\
\sigma^\mathrm{SI}_{\mathcal{L}_{17}} &=& \sigma_0 \left( \frac{g_D^2}{16} \right)^2 \left( 1 + \frac{2}{3} \ln \frac{m_\mu^2}{M^2} \right)^2, \label{eq:DDL17}  \\
\sigma^\mathrm{SI}_{\mathcal{L}_{18}} &=& \sigma_0 \left( \frac{g_D^2}{12} \right)^2 \left( \ln \frac{m_\mu^2}{M^2} \right)^2, \label{eq:DDL18} \\
\sigma^\mathrm{SI}_{\mathcal{L}_{22}} &=& \sigma_0 \left( \frac{g_D^2}{12} \right)^2 \left( \ln \frac{m_\mu^2}{M^2} \right)^2, \label{eq:DDL22}
\end{eqnarray}
where $\alpha$, $m_N$, $Z$, $A$ and $m_\mu$ are fine structure constant, target nucleus's mass, target nucleus's charge, target nucleus's mass number and muon's mass respectively. 
The velocity of DM near the earth is $v \sim 10^{-3}$, reduced mass of DM-nucleus system is $\mu_N = \frac{m_N m_D}{m_N + m_D}$, 
and the cut-off scale assumed to be $\Lambda = \frac{M}{\sqrt{g_D g_f}}$. The coefficient $\sigma_0 = \frac{\alpha^2 Z^2 \mu_N^2}{\pi^3 A^2 M^4}$.

\section{Muon $g-2$ formulas for models with $Z_2$ even mediators}\label{append:gm2}

The predictions of muon $g-2$ excess from various simplified muonphilic DM models are based on their mediator types~\cite{Agrawal:2014ufa,Queiroz:2014zfa,Chen:2015vqy,Kowalska:2017iqv,Calibbi:2018rzv,Abu-Ajamieh:2018ciu}.
In this appendix, we provide the exact expressions of additional muon $g-2$ contribution $\delta{a_\mu}$ 
to the standard model prediction as the follows:  
\begin{itemize}
\item For the vector mediator, \textit{i.e.} $\mathcal{L}_{5,7,11,15}$, we obtain 
\begin{equation}
\delta{a_\mu} = \frac{g_f^2}{8\pi^2}\int_0^1\frac{2m^2_\mu z (1-z)^2}{m^2_\mu (1-z)^2 + M^2 z}\, dz.
\label{eq:10}
\end{equation}

\item For the axial-vector mediator $\mathcal{L}_{6,8,12,16}$, we obtain  
\begin{equation}
\delta{a_\mu} = - \frac{g_f^2}{4\pi^2}\frac{m^2_\mu}{M^2}\int_0^1\frac{2m^2_\mu (1-z)^3+z(1-z)(3+z)M^2}{m^2_\mu (1-z)^2 + M^2 z}\, dz.
\label{eq:11}
\end{equation}

\item For the scalar mediator $\mathcal{L}_{1,3,9,13}$, we obtain 
\begin{equation}
\delta{a_\mu} = \frac{g_f^2}{8\pi^2}\frac{m^2_\mu}{M^2}\int_0^1\frac{M^2 (1+z)(1-z)^2}{m^2_\mu (1-z)^2 + M^2 z}\, dz.
\label{eq:12}
\end{equation}

\item For the pseudo-scalar mediator $\mathcal{L}_{2,4,10,14}$, we obtain 
\begin{equation}
\delta{a_\mu} = - \frac{g_f^2}{8\pi^2}\frac{m^2_\mu}{M^2}\int_0^1\frac{M^2 (1-z)^3}{m^2_\mu (1-z)^2 + M^2 z}\, dz.
\label{eq:13}
\end{equation}

\end{itemize}

\end{document}